\documentclass[onecollarge]{svjour2}       
\smartqed  
\usepackage[dvipdfmx]{graphicx}
\usepackage{amsmath,amssymb}
\usepackage{mathrsfs}
\usepackage{dcolumn}
\usepackage{bm}
\usepackage{tabularx}

\newcommand{\mc}{\mathcal}
\newcommand{\tl}{\tilde}
\newcommand{\ra}{\rangle}
\newcommand{\la}{\langle}
\newcommand{\ve}{\varepsilon}
\newcommand{\tmV}{\tl{\mc{V}}}
\newcommand{\om}{\omega}
\newcommand{\Om}{\Omega}
\newcommand{\sgn}{{\rm \>sgn}}
\newcommand{\tb}{\textbf}

\newcommand{\hv}{\hat{v}}

\newcommand{\hmV}{\hat{\mathcal{V}}}

\newcommand{\hxi}{\hat{\xi}}

\def\XXint#1#2#3{{\setbox0=\hbox{$#1{#2#3}{\int}$}
\vcenter{\hbox{$#2#3$}}\kern-.5\wd0}}

\journalname{Journal of Statistical Physics}
\begin{document}

\title{Asymptotic derivation of Langevin-like equation with non-Gaussian noise and its analytical solution}


\author{Kiyoshi Kanazawa \and Tomohiko G. Sano \and Takahiro Sagawa \and Hisao Hayakawa}


\institute{
			K. Kanazawa \and T. G. Sano \and H. Hayakawa	\at
				Yukawa Institute for Theoretical Physics, Kyoto University, Kitashirakawa-oiwake cho, Sakyo-ku, Kyoto 606-8502, Japan\\
				\email{kiyoshi@yukawa.kyoto-u.ac.jp}
			\and
			T. Sagawa	\at
				Department of Basic Science, The University of Tokyo, Komaba, Meguro-ku, 153-8902, Japan		
}

\date{Received: date / Accepted: date}

\maketitle

\begin{abstract}
	We asymptotically derive a non-linear Langevin-like equation with non-Gaussian white noise for a wide class of stochastic systems associated with multiple stochastic environments, 
	by developing the expansion method in our previous paper [K.~Kanazawa {\it et al.}, arXiv: 1407.5267 (2014)]. 
	We further obtain a full-order asymptotic formula of the steady distribution function in terms of a large friction coefficient for a non-Gaussian Langevin equation with an arbitrary non-linear frictional force. 
	The first-order truncation of our formula leads to the independent-kick model and the higher-order correction terms directly correspond to the multiple-kicks effect during relaxation. 
	We introduce a diagrammatic representation to illustrate the physical meaning of the high-order correction terms. 
	As a demonstration, we apply our formula to a granular motor under Coulombic friction and get good agreement with our numerical simulations. 
	
\end{abstract}

\section{Introduction}
	Stochastic theory has been a powerful tool to understand phenomena in various fields, 
	such as physics~\cite{Kubo}, chemistry~\cite{Prigogine}, biophysics~\cite{Biophysics}, and economics~\cite{Econophysics}. 
	In particular, the Langevin model with the white Gaussian noise is often used in modeling fluctuating systems~\cite{Langevin}. 
	Its microscopic foundation has been understood for a system driven by a single stochastic environment in terms of microscopic theories~\cite{vanKampen,vanKampenB,Zwanzig,SekimotoZwanzig}.
	For example, van Kampen's theory~\cite{vanKampen,vanKampenB} predicts that a stochastic system associated with a single environment is asymptotically described by a Gaussian model in the large system size limit (or equivalently, the small noise limit).
	Furthermore, the Gaussian Langevin model is sufficiently simple to be analytically solvable for a wide class of setups~\cite{Gardiner}.
	For these reasons, the Gaussian Langevin model has been accepted as a minimal model for the Brownian motion with a single environment, 
	and has played an important role in the recent development of thermodynamics of small systems~\cite{Bustamante,Liphardt,Trepagnier,Blickle,Garnier,Ciliberto,Sekimoto1,Sekimoto2,SekimotoB,SeifertR1,SeifertR2,Evans,Gallavotti,Jarzynski,Crooks,Seifert,Kurchan}. 

	On the other hand, stochastic systems associated with multiple environments have not been fully understood. 
	The role of multiple stochastic environments is significant for athermal systems, 
	where both thermal and athermal fluctuations coexist because of external energy injection from the reservoirs. 
	For example, athermal noise (e.g., avalanche~\cite{Gabelli,Zaklikiewicz} or shot noise~\cite{Blanter}) plays an important role as well as thermal noise in electrical circuits. 
	In granular and biological systems, it is known that the granular noise~{\cite{Eshuis,Gnoli1,Gnoli2,Gnoli3}} and active noise~{\cite{Ben-Isaac,Toyota}}, respectively, appear because of external vibration and consumption of adenosine triphosphate (ATP). 
	These systems cannot be addressed by the conventional microscopic theories because they are coupled with multiple environments. 
	A generalization of van Kampen's approach toward athermal systems has recently been formulated in Ref.~\cite{Kanazawa1} by considering systems associated with two different environments, i.e., thermal and athermal environments. 
	In Ref.~\cite{Kanazawa1}, it is predicted that athermal stochastic systems are universally characterized by Langevin-like equations driven by non-Gaussian noise, 
	which is consistent with experimental reports on athermal fluctuations in electric, granular, and biological systems~\cite{Gabelli,Gnoli3,Ben-Isaac,Toyota}. 
	Such non-Gaussian models are expected to be important in non-equilibrium statistical mechanics for athermal systems~\cite{Kanazawa1,Luczka,BauleCohen,Kanazawa2,Morgado,Kanazawa3,Kanazawa4}.
	
	In this paper, we extend the formulation in Ref.~\cite{Kanazawa1} to non-linear frictional systems. 
	We asymptotically derive a non-linear Langevin-like equation with non-Gaussian noise in the small noise limit for the environments. 
	We further obtain an analytic solution for an arbitrary non-Gaussian Langevin equation with a non-linear frictional force. 
	We derive a full-order asymptotic formula in terms of a large frictional coefficient for the velocity distribution function (VDF), 
	and show that the first-order approximation corresponds to the independent-kick model, which was phenomenologically introduced in Ref.~\cite{Talbot}. 
	We also show that the higher-order terms directly correspond to the multiple-kicks effect during relaxation, and introduce a diagrammatic representation to illustrate the higher-order terms.
	As a demonstration, we address the stochastic motion of a granular motor under dry friction to verify the validity of our theory. 
	
	This paper is organized as follows: 
	In Sec.2, we asymptotically derive the non-Gaussian Langevin equation with a non-linear friction by a small noise expansion. 
	In Sec.3, we study the steady distribution function of the non-Gaussian Langevin equation, and derive the full-order asymptotic solution in terms of the inverse of the frictional coefficient. 
	In Sec.4, we study a granular motor under dry friction and verify our formulation numerically. 
	In Appendix.~{\ref{app:gmotor_viscous}}, we apply our formulation to the nonequilibrium steady state of a rotor in granular and molecular gases. 
	In Appendix.~{\ref{app:CN:IntegralEq}}, we derive the solution of the iterative integral equation for the Fourier representation of the distribution. 
	In Appendix.~{\ref{app:CN:tail_cubic}}, we check the asymptotic tail of the Fourier representation of the distribution for the cubic friction.
	In Appendix.~{\ref{app:CN:renormalized}}, we check the validity of the first-order renormalized solution for the cubic friction. 
	In Appendix.~{\ref{app:CN_cumulant_g}}, we show the detailed derivation of the cumulant function for the granular noise. 
	In Appendix.~{\ref{app:CN_first_gm}}, we show the detailed derivation of the first-order formula of the steady distribution function for the granular motor.

\section{Asymptotic derivation of non-Gaussian Langevin equations}
	\subsection{Setup}
\begin{figure}
			\centering
			\includegraphics[width=140mm]{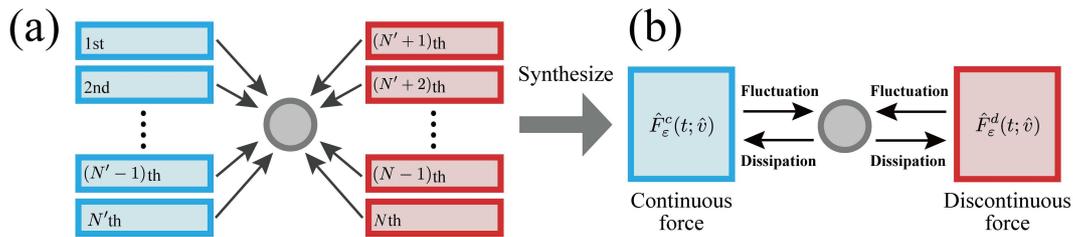}
			\caption{	
						(Color online) Schematics of the setup of a Brownian particle attached to multiple environments. 
						(a) We assume that all environments are classified into two types: one is the continuous force that consists of deterministic force and the stochastic Gaussian noise, and the other is the discontinuous force that describes jump processes. 
						(b) We synthesize the environments into two environments. 
					}
			\label{fig:sc_multi_env}
		\end{figure}

		Let us consider a Brownian particle moving in one dimensional space coupled with multiple environments (see Fig.~{\ref{fig:sc_multi_env}(a)}). 
		For simplicity, we assume that the mass of the particle is unity and that its motion obeys the Markovian dynamics characterized by a small parameter $\ve$. 
		As will be illustrated later, $\ve$ characterizes the amplitude of noise terms and corresponds to the inverse of the system size as in Refs.~\cite{vanKampen,vanKampenB}.
		The dynamics of the velocity $\hat v$ of the particle then obey the following master equation (so-called the differential Chapman-Kolmogorov equation~\cite{Gardiner}): 
		\begin{equation}
			\frac{\partial P(v,t)}{\partial t} = \sum_{i=1}^N L_{i;\ve} P(v,t),\label{eq:ME_general}
		\end{equation}
		where $P(v,t)\equiv P(\hat v(t)=v)$ is probability density, $N$ is the number of stochastic environments, and $L_{i;\ve}$ is the time-independent Liouville operator originating from the $i$th environment. 
		Throughout this paper, we denote a stochastic variable by a variable with a hat such as $\hat{v}$.
		For $0\leq i \leq N'<N$ with an integer $N'$, we assume that the $i$th environment frequently interacts with the tracer particle and is described by a continuous force 
		(the combination of the deterministic force and the Gaussian noise): 
		\begin{equation}
			L_{i;\ve}P(v,t) = \left[\frac{\partial }{\partial v}\alpha_{i;\ve}(v) + \frac{1}{2}\frac{\partial^2 }{\partial v^2}\beta^2_{i;\ve}(v)\right]P(v,t), 
		\end{equation}
		where $\alpha_{i;\ve}(v)$ is deterministic friction and $\beta_{i;\ve}^2(v)$ is the variance of the Gaussian noise. 
		Note that these operators have locality, which describe diffusion processes. 
		In fact, the sample paths related to $L_{i;\ve}$ for $1\leq i \leq N'$ are continuous but not differentiable almost everywhere~\cite{Gardiner}.
		For $N'< i \leq N$, we assume that the $i$th environment rarely but strongly interacts with the tracer particle and is described by the Markovian jump process: 
		\begin{equation}
			L_{i;\ve}P(v,t) = \int_{-\infty}^\infty dy\left[P(v-y,t)T_{i;\ve}(v-y;y) - P(v,t)T_{i;\ve}(v;y)\right],
		\end{equation}
		where $T_{i;\ve}(v;y)$ is the transition rate from $v$ with velocity jump $y$, the first term on the right-hand side (rhs) represents the probability inflow into $v$, and the second term represents the probability outflow from $v$. 
		Note that these operators have non-locality because they describe non-local jump processes.
		\begin{figure}
			\centering
			\includegraphics[width=150mm]{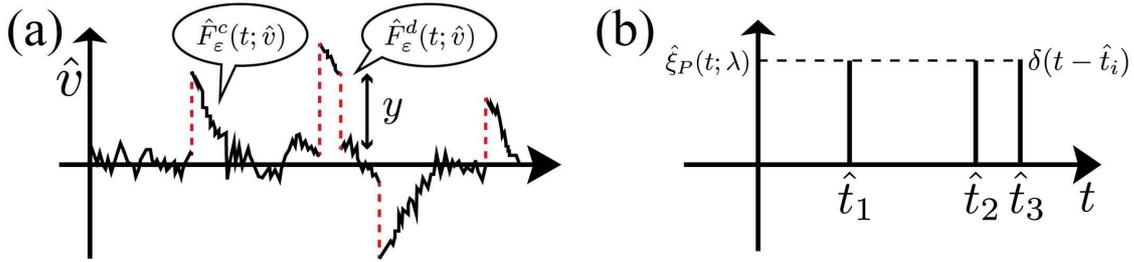}
			\caption{
						(Color online)
						(a) A typical trajectory of the system obeying the master equation~{(\ref{eq:ME_general})}. 
						The continuous motion (the solid line) is induced by the continuous force $\hat F_\ve^c (t;\hat v)$ characterized by the continuous Liouville operator $\mc{L}_\ve^c$. 
						$\mc{L}_\ve^c$ is characterized by the friction function $A_\ve(\hat v)$ and the multiplicative Gaussian noise $B_\ve(\hat v)\cdot \hat \xi_G$. 
						The discontinuous motion (the dashed line) is induced by the discontinuous force $\hat F_\ve^d(t;\hat v)$ characterized by the discontinuous Liouville operator $\mc{L}_\ve^d$.
						$\mc{L}_\ve^d$ is characterized by the transition rate $W_\ve (v;y)$ with the flight distance $y$.
						(b) A typical trajectory of the Poisson noise $\hat \xi_P(t;\lambda)$. 
						The Poisson noise is a $\delta$-type singular noise, and the systems driven by the Poisson noise move discontinuously like Fig.~{\ref{fig:CD_trajectory}(a)}. 
					}
			\label{fig:CD_trajectory}
		\end{figure}
		We assume that $T_{i;\ve}(v;y)$  converges to zero for $y\to\pm \infty$ in a sufficiently rapid speed (e.g., $T_{i;\ve}(v^*_{\ve};y)\lesssim e^{-|y|/y^*_{i;\ve}}$ for $y\to\infty$, where $v^*_{\ve}$ is the typical velocity scale and $y^*_{i;\ve}$ is the typical velocity jump scale). 
		We then introduce the following synthesized Liouville operators (see Fig.~{\ref{fig:sc_multi_env}}(b)):
		\begin{align}
			\mc{L}^c_{\ve} P(v,t) &= \left[\frac{\partial }{\partial v}A_{\ve}(v) + \frac{1}{2}\frac{\partial^2 }{\partial v^2}B^2_{\ve}(v)\right]P(v,t),\label{eq:ME_FPpart}\\
			\mc{L}^d_{\ve} P(v,t) &= \int_{-\infty}^\infty dy\left[P(v-y,t)W_{\ve}(v-y;y) - P(v,t)W_{\ve}(v;y)\right],\label{eq:ME_Jumppart}
 		\end{align}
 		where $A_{\ve}(v)\equiv \sum_{i=1}^{N'} \alpha_{i;\ve}(v)$, $B^2_{\ve}(v)\equiv \sum_{i=1}^{N'}\beta_{i;\ve}^2(v)$, and $W_{\ve}(v;y)\equiv \sum_{i=N'+1}^N T_{i;\ve}(v;y)$. 
		The Liouville operators $\mc{L}^c_{\ve}$ and $\mc{L}^d_{\ve}$ describe continuous and discontinuous motions induced by stochastic forces, respectively (Fig.~{\ref{fig:CD_trajectory}(a)}). 
		By introducing the white Gaussian noise $\hat \xi_G(t)$ satisfying $\la\hxi_G(t)\ra=0$ and $\la \hat \xi_G(t)\hat \xi_G(s)\ra=\delta(t-s)$ and the Poisson noise $\hat \xi_P(t;\lambda)$ with transition rate $\lambda$, 
		Eqs.~{(\ref{eq:ME_general})}, {(\ref{eq:ME_FPpart})}, and {(\ref{eq:ME_Jumppart})} are equivalent to the stochastic differential equation: 
		\begin{equation}
			\frac{d\hat v}{dt} = \hat F^c_{\ve}(t;\hat v) + \hat F^d_{\ve}(t;\hat v)
		\end{equation}
		with 
		\begin{align}
			\hat F^c_{\ve}(t;\hat v) &\equiv -A_{\ve}(\hat v) + B_{\ve}(\hat v)\cdot \xi_G(t),\\
			\hat F^d_{\ve}(t;\hat v) &\equiv \sum_{y} y \hat \xi_P(t;\lambda_{y}^{\ve}(\hat v)), \label{eq:F^d_ve}
		\end{align}
		where the symbol $\sum_y$ takes the summation for velocity jump $y$, and we introduce conditional transition rate $\lambda_{y}^{\ve}(\hat v) \equiv dyW_{\ve}(\hat v;y)$. 
		In the following, we denote the ensemble averages of stochastic quantities as $\la \hat A\ra$. 
		We here stress that the fluctuation terms $\hat F^c_\ve(t;\hat v)$ and $\hat F^d_\ve(t;\hat v)$ have correlation with the velocity of the tracer $\hat v$, 
		which implies that the environmental fluctuation is not white noise but complicated stochastic force. 
		We also note that the Poisson noise is the sum of $\delta$-type spike noise terms (Fig.~{\ref{fig:CD_trajectory}(b)}) as 
		\begin{equation}
			\hat \xi_P(t;\lambda) = \sum_{i=1}^\infty \delta(t-\hat t_i),
		\end{equation}
		where $\{\hat t_i\}_i$ are the times at which the Poisson flights happen and are characterized by the transition rate $\lambda$. 
		The transition rate $\lambda$ characterizes the typical interval between two successive Poisson flights as $\hat t_{i+1}-\hat t_i \sim \lambda^{-1}$.
		We also note that the summation in Eq.~{(\ref{eq:F^d_ve})} can be formally written as the integral form: $\hat F^d_{\ve}(t;\hat v) = \int_{-\infty}^{\infty} y\hat \xi_P(t;\lambda_{y}^{\ve}(\hat v))$~\cite{Levy}. 
		
	\subsection{Derivation of non-Gaussian Langevin equations with non-linear friction terms}\label{sec:NGL_nonliner}
\begin{figure}
			\centering
			\includegraphics[height=28mm]{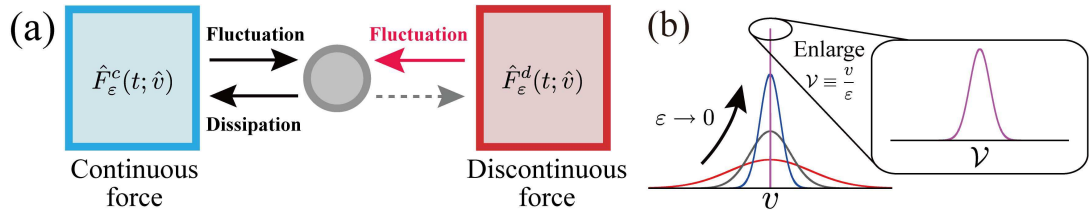}
			\caption{
						(Color online)
						(a) Schematic of the strong friction assumption (ii). 
						The frictional effect in the continuous force $\hat F_\ve^c(t;\hat v)$ is more relevant than that in the discontinuous force $\hat F_\ve^d(t;\hat v)$. 
						We stress that the discontinuous force $\hat F_\ve^d(t;\hat v)$ is only relevant to the fluctuation to the system (the red arrow in the figure). 
						(b) Schematic of the divergence of $P(v,t)$ to the $\delta$-type distribution $\delta(v)$ in the limit $\ve\to 0$. 
						By introducing a scaled variable $\mc{V}\equiv v/\ve$, the peak of the distribution is enlarged, where the Langevin-like description is valid. 
					}
			\label{fig:Assumption1}
		\end{figure}

		In this subsection, we derive non-Gaussian Langevin equations with non-linear friction terms for more general setups than those in Ref.~\cite{Kanazawa1}. 
		Non-linear frictions are ubiquitous in nature~\cite{Persson2000,Wang2009,Wang2012} and are known to appear in systems such as granular~\cite{Kawamura,Olsson,Jop2006}, biological~\cite{Bormuth,Veigel,Jagota2011} and atomic-surface ones~\cite{Urbash2004,Li2011,Weymouth2013}.
		We note that non-linear frictions can be discontinuous functions with respect to velocity in general (e.g., Coulombic friction), 
		and their singular effects on stochastic properties have been interesting topics~\cite{Eshuis,Gnoli1,Gnoli2,Gnoli3,Kawarada,Hayakawa,deGenne,Touchette2,Menzel,Baule,Talbot2,Touchette,Sarracino,Baule2,SanoHayakawa}. 
		Indeed, as will be shown in the next section, the distribution function can be strongly singular around the peak. 
		We here introduce critical assumptions as follows:
		\begin{quote}
		(i) {\it {Small noise assumption}}: The noise amplitudes in $\hat F^c_\ve (t;\hat v)$ and $\hat F^d_\ve (t;\hat v)$ are small. 
				In other words, their stochastic parts are scaled by a small positive constant $\ve$ as 
				\begin{align}
					b_{\ve}(\hat v) \cdot \xi_G(t) &= \ve \mc{B}(\hat v)\cdot \hat \xi_G(t), \label{eq:epsilon_continuous}\\
					\hat F^d_{\ve}(t;\hat v) 		 &= \ve \hat \eta (t;\hat v), \label{eq:epsilon_jump}
				\end{align}
				where $\mc{B}(\hat v)$ is a non-negative smooth function independent of $\ve$ 
				and $\hat \eta(t;\hat v)$ is a Markovian jump force whose transition rate $\overline{W}(v;\mc{Y})$ (i.e., the Poisson jump rate with the jump amplitude $\mc{Y}$ on the condition $\hat v(t)=v$) is independent of $\ve$ and is a smooth function in terms of $v$. 
				We note that $\hat \eta(t;\hat v)$ can be decomposed into the following form: 
				\begin{equation}
					\hat \eta(t;\hat v) = \sum_{\mc{Y}}\mc{Y}\hat \xi_P(t;\tl{\lambda}_{\mc{Y}}(\hat v)),
				\end{equation}
				where $\tl{\lambda}_{\mc{Y}}(\hat v)\equiv d\mc{Y}\overline{W}(\hat v;\mc{Y})$. 
				We here stress that $\hat \eta(t;\hat v)$ is independent of $\ve$, corresponding to the $\ve$-independence of $\overline{W}(v;\mc{Y})$. 
				
		(ii) {\it {Strong deterministic friction}}: The friction function $A_{\ve}(\hv)$ can be expanded as, 
				\begin{equation}
					A_{\ve}(\hv) = \sum_{n=1}^\infty \frac{\ve^n}{n!} \mc{A}_{(n)}(\hmV),\label{eq:assump_NL}
				\end{equation}
				where $\hmV\equiv \hv/\ve$ is the scaled velocity and $\mc{A}_{(n)}(\hmV)$ is independent of $\ve$.
				This scaling implies that the frictional effect in $\hat F_\ve^d(t;\hat v)$ is negligible compared with that in $\hat F_\ve^c(t;\hat v)$ (see Fig.~{\ref{fig:Assumption1}(a)} as a schematics).
				
		(iii) {\it {Stable deterministic friction}}: Both $A_{\ve}(\hv)$ and $\mc{A}_{(1)}(\hmV)$ are piecewise smooth functions of $\hv$ and $\hmV$, and have the single stable zero points $\hv=\hmV=0$ as 
				\begin{align}
					A_{\ve}(0)=0, \>\>\> &A_{\ve}(\hv) >0, \>\>\>A_{\ve}(-\hv) <0,\\
					\mc{A}_{(1)}(0)=0, \>\>\> &\mc{A}_{(1)}(\hmV) >0, \>\>\>\mc{A}_{(1)}(-\hmV) <0,
				\end{align}
				where $\hv$ and $\hmV$ are arbitrary positive numbers. 
		\end{quote}
		We note that the condition (i) is the weak-coupling condition between the system and the environment,
		which is necessary to truncate the environmental correlation. 
		We also note that the scalings~{(\ref{eq:epsilon_continuous})} and {(\ref{eq:epsilon_jump})} are equivalent to  
		\begin{align}
			B^2_{\ve}(v) &= \ve^2 \mc{B}^2(v),\\
			W_{\ve}(v;y) &=\frac{1}{\ve}\overline{W}\left(v;\frac{y}{\ve}\right),\label{eq:scaling_TR}
		\end{align}
		where $\overline{W}(v;\mc{Y})$ is the scaled jump rate independent of $\ve$ with the scaled jump $\mc{Y}\equiv y/\ve$. 
		The scaling~{(\ref{eq:scaling_TR})} can be derived as follows: 
		According to the scaling~{(\ref{eq:epsilon_jump})}, the jump size $y$ by the discontinuous force $\hat F^d_{\ve}$ should be scaled as $\mc{Y}\equiv y/\ve$ to remove the $\ve$-dependence. 
		Then, the following relation holds:
		\begin{equation}
			dyW_{\ve}(v;y) = d\mc{Y}\overline{W}(v;\mc{Y}),
		\end{equation}
		which implies the scaling~{(\ref{eq:scaling_TR})}. 
		The scaling~{(\ref{eq:scaling_TR})} is essentially equivalent to that introduced by van Kampen~\cite{vanKampen,vanKampenB}, where $\ve$ corresponds to the inverse of the system size. 
		We also note two examples satisfying the assumptions (ii) and (iii): 
		The first example is the viscous friction $A_{\ve}(\hv)=\gamma \hv$ with an $\ve$-independent parameter $\gamma>0$. 
		The second example is Coulombic friction $A_{\ve}(\hv)=\ve \gamma \sgn(\hv)$ with an $\ve$-independent parameter $\gamma>0$. 
		We note that Coulombic friction appears for systems in contact with solid~\cite{Kawamura,Olsson,Talbot}. 
		We also note that the sign function ${\rm sgn}(v)$ is defined as follows: 
		For $v>0$, ${\rm sgn}(v) = +1$. For $v=0$, ${\rm sgn}(v) = 0$. For $v<0$, ${\rm sgn}(v) = -1$.

		We next derive the non-Gaussian Langevin equation using an asymptotic expansion in terms of $\ve$. 
		In the small noise limit $\ve \to 0$, the steady distribution function $P_{\rm SS}(v)\equiv \lim_{t\to \infty}P(v,t)$ converges to the $\delta$-function around the stable point $v=0$ as $\lim_{\ve\to 0}P_{\rm SS}(v)=\delta(v)$, 
		because the small noise expansion is a singular perturbation~\cite{SingularPerturbation} (see Fig.~{\ref{fig:Assumption1}(b)}).
		In order to solve this singular perturbation, we have to introduce an appropriate scaled variable 
		\begin{equation}
			\mc{V} \equiv \frac{v}{\ve}, \label{eq:NGL_scaling}
		\end{equation}
		which enlarges the peak of the distribution $P(v,t)$, where the Langevin-like description is asymptotically valid (see Fig.~{\ref{fig:Assumption1}}(b)). 
		
		On the basis of the above assumptions (i), (ii) and (iii), let us derive non-linear Langevin equations. 
		By introducing the scaled variable $\mc{V}=v/\ve$ and the scaled distribution $\mc{P}(\mc{V},t)=\ve P(v,t)$, the master equation~{(\ref{eq:ME_general})} can be written as 
		\begin{align}
			\frac{\partial \mc{P}(\mc{V},t)}{\partial t} &= \sum_{n=0}^\infty \frac{\ve^n}{n!}	\Bigg[\bigg\{\frac{\partial }{\partial \mc{V}}\frac{\mc{A}_{(n+1)}(\mc{V})}{n+1}+\frac{\mc{B}_{(n)}^{2*}}{2}\frac{\partial^2 }{\partial \mc{V}^2}\mc{V}^n\bigg\}P(\mc{V},t) \notag\\
																									&+ \int_{-\infty}^\infty d\mc{Y}\overline{W}_{(n)}^*(\mc{Y})\bigg\{\mc{P}(\mc{V-Y},t)(\mc{V-Y})^n-\mc{P}(\mc{V},t)\mc{V}^n\bigg\} \Bigg],\label{eq:SME_NL_tran}
		\end{align}
		where we have used Eq.{(\ref{eq:assump_NL})} and  
		\begin{equation}
			\mc{B}^2(\ve \mc{V})			= \sum_{n=0}^\infty \frac{\ve^n}{n!}\mc{V}^n\mc{B}_{(n)}^{2*},\>\>\>
			\overline{W}(\ve\mc{V};\mc{Y})	= \sum_{n=0}^\infty \frac{\ve^n}{n!}\mc{V}^n\overline{W}_{(n)}^*(\mc{Y}). 
		\end{equation}
		We then obtain the following reduced master equation in the limit $\epsilon\to 0$ 
		\begin{equation}
			\frac{\partial \mc{P}(\mc{V},t)}{\partial t}	= \Bigg[\bigg\{\frac{\partial }{\partial \mc{V}}F(\mc{V})+\frac{\sigma^2}{2}\frac{\partial^2 }{\partial \mc{V}^2}\bigg\}P(\mc{V},t)
															+ \int_{-\infty}^\infty d\mc{Y}\mc{W}(\mc{Y})\bigg\{\mc{P}(\mc{V-Y},t)-\mc{P}(\mc{V},t)\bigg\} \Bigg], \label{eq:SME_NL}
		\end{equation}
		where we have introduced the friction function $F(\mc{V})\equiv \mc{A}_{(1)}(\mc{V})$, the variance of the Gaussian noise $\sigma^2\equiv \mc{B}_{(0)}^{2*}$, and the transition rate $\mc{W}(\mc{Y})\equiv \overline{W}^*_{(0)}(\mc{Y})$. 
		Note that the transition rate $\mc{W}(\mc{Y})$ is independent of $\mc{V}$, which implies that the environmental correlation disappears and the discontinuous stochastic force is reduced to white noise. 
		Equation~{(\ref{eq:SME_NL})} is then equivalent to the non-linear non-Gaussian Langevin equation:
		\begin{equation}
			\frac{d\hmV}{dt} = -F(\hmV) + \sigma \hxi_{\rm G}+ \hxi_{\rm NG},\label{eq:NGL_NL_derived}
		\end{equation}
		with the white non-Gaussian noise $\hat \xi_{\rm NG}$ whose transition rate is given by $\mc{W}(\mc{Y})$. 
		We note that the frictional effect only appears from the continuous force $\hat F_\ve^c(t;\hv)$, not from the discontinuous force $\hat F_\ve^d(t;\hv)$ (see Fig.~{\ref{fig:Assumption1}}(a)).

	\subsection{Weak friction cases: Reduction to the Gaussian Langevin equation}
		\begin{figure}
			\centering
			\includegraphics[height=25mm]{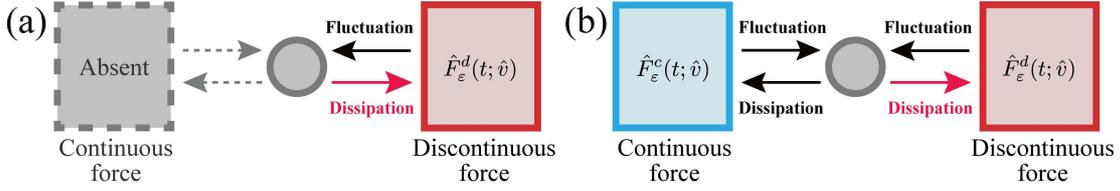}
			\caption{
						(Color online)
						(a) Schematic of the original setup addressed by van Kampen~{\cite{vanKampen,vanKampenB}}, where the continuous force is absent as $\hat F^c_{\ve}(t;\hat v)=0$.
						(b) Schematic of the assumption (ii'), where the frictional effect in the continuous force $\hat F_\ve^c(t;\hat v)$ is comparable with that in the discontinuous force $\hat F_\ve^d(t;\hat v)$ as seen in the existence of the red arrow. 
					}
			\label{fig:Assumption3}
		\end{figure}
\begin{figure}
			\centering
			\includegraphics[width=130mm]{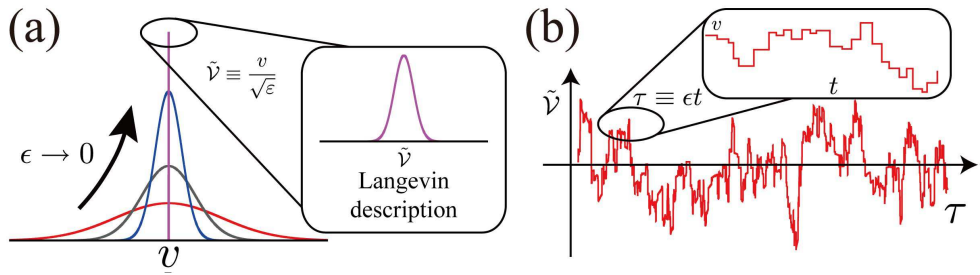}
			\caption	{
								(Color online)
								(a) Schematic of the divergence of the distribution $P(v,t)$ in the limit $\ve\to0$.
								By introducing a scaled velocity $\tl{\mc{V}}\equiv v/\sqrt{\ve}$, the peak of the distribution is appropriately enlarged. 
								(b) Schematic of a typical trajectory of the system. 
								By introducing a scaled time $\tau\equiv \ve t$, the trajectory is appropriately coarse-grained in terms of time. 
							}
			\label{fig:Scaling_G}
		\end{figure}

		We next analyze the case that the friction $A_\ve(\hat v)$ is weak or absent. 
		We note that the original setup by van Kampen is the case without the continuous force: $\hat F_{\ve}^c(t;\hat v)=0$ (see Fig.~{\ref{fig:Assumption3}(a)}).
		We make the following assumptions (ii')-(iv') instead of the above assumptions {(ii)-(iii)}: 
		\begin{quote}
			(ii') {\it {Weak deterministic friction}}: The friction $A_{\ve}(\hv)$ is scaled by $\ve$ as 
				\begin{equation}
					A_{\ve}(\hv) = \ve \mc{A}(\hv),
				\end{equation}
				where $\mc{A}(\hv)$ is independent of $\ve$. 
				This scaling implies that the frictional effect in $\hat F_\ve^d(t;\hat v)$ is comparable with that in $\hat F_\ve^c(t;\hat v)$ (see Fig.~{\ref{fig:Assumption3}(b)} as a schematic).

			(iii') {\it {Stable deterministic friction}}: The friction $\mc{A}(\hv)$ is zero ($\mc{A}(\hv)=0$), or is a smooth function of $\hv$ which has a single stable zero point at $\hv=0$ as 
				\begin{equation}
					\mc{A}(0)=0,  \>\>\> \mc{A}'(0)\equiv (d\mc{A}/d\hv)|_{\hv=0}>0. 
				\end{equation}
						
			(iv') {\it {Stable jump force}}: 
				The jump force $\hat F_\ve^d(t;\hv)$ is stable around $\hv=0$. 
				In other words, the following relations are assumed for the jump rate: 
				Let us introduce the scaled variable $\mc{Y}\equiv y/\epsilon$ and the scaled Kramers-Moyal coefficients 
				\begin{equation}
					\mc{K}_n(v) \equiv (-1)^n\int_{-\infty}^\infty d\mc{Y}\mc{Y}^n \overline{W}(v;\mc{Y}). 
				\end{equation}
				We assume that the Kramers-Moyal coefficients $\{\mc{K}_n(v)\}_{n\geq 1}$ are smooth functions and 
				the first-order coefficient $\mc{K}_1(v)$ has a single stable zero point $v=0$ as
				\begin{equation}
					\mc{K}_1(0)=0, \>\>\> \mc{K}_1' (0) \equiv (d\mc{K}_1/dv)|_{v=0} >0. 
				\end{equation}
		\end{quote}
		Under the assumptions (i) and (ii')-(iv'), we derive a Gaussian Langevin equation. 
		According to the Kramers-Moyal expansion, we obtain 
		\begin{equation}
			\mc{L}_\ve^d P(v,t) = \sum_{n=1}^\infty \frac{\ve^n}{n!}\frac{\partial^n }{\partial v^n}[\mc{K}_n(v)P(v,t)]. 
		\end{equation}
		Then, the master equation~{(\ref{eq:ME_general})} can be written as 
		\begin{equation}
			\frac{\partial P(v,t)}{\partial t} = \left[\ve\frac{\partial }{\partial v}\mc{A}(v) + \frac{\ve^2}{2}\frac{\partial^2 }{\partial v^2}\mc{B}^2(v)\right]P(v,t) + \sum_{n=1}^\infty \frac{\ve^n}{n!}\frac{\partial^n }{\partial v^n}\left[\mc{K}_n(v)P(v,t)\right].
		\end{equation}
		We here introduce the following scaled variables: 
		\begin{equation}
			\tl{\mc{V}}\equiv\frac{v}{\sqrt{\ve}}, \>\>\> \tau \equiv \ve t,\label{eq:GL_scaling}
		\end{equation}
		where the scaled velocity $\tilde{\mc{V}}$ is introduced to enlarge the peak of the distribution (see Fig.~\ref{fig:Scaling_G}(a)), 
		and the scaled time $\tau$ is introduced to describe the coarse-grained dynamics (see Fig.~\ref{fig:Scaling_G}(b)). 
		Note that the appropriate scaled variables~{(\ref{eq:GL_scaling})} are different from the scaled variable~{(\ref{eq:NGL_scaling})} in Sec.\ref{sec:NGL_nonliner}.
		This difference is important because the introduction of appropriate scaled variables is the key to the singular perturbation. 
		We then obtain the Kramers-Moyal expansion for the scaled distribution $\mc{P}(\tmV,\tau)\equiv \sqrt{\ve}P(v,t)$ as 
		\begin{align}
			\frac{\partial \mc{P}(\tl{\mc{V}},t)}{\partial \tau} 	=& \sum_{m=0}^{\infty}\ve^{m/2} \left[\frac{\partial }{\partial \tmV}\frac{\tmV^{m+1}}{(m+1)!}\left(A_{(m+1)}^*+K_{1;(m+1)}^*\right) + \frac{1}{2}\frac{\partial^2 }{\partial \tmV^2}\frac{\tmV^m}{m!}\left(\mc{B}_{(m)}^{2*}+K_{2;(m)}^*\right)\right]\mc{P}(\tmV,t) \notag\\
																	&+ \sum_{n=3}^\infty\sum_{m=0}^\infty \frac{\ve^{(n+m-2)/2}}{n!m!}\mc{K}_{n;(m)}^*\frac{\partial^n }{\partial \tmV^n}\left[\tmV^m \mc{P}(\tmV,t)\right],
		\end{align}
		where we expand 
		\begin{equation}
			\mc{A}(\ve^{1/2}\tmV) = \sum_{m=1}^\infty \frac{\ve^{m/2}\tmV^m}{m!}\mc{A}^*_{(m)},\>\>\>
			\mc{B}^2(\ve^{1/2}\tmV) = \sum_{m=0}^\infty \frac{\ve^{m/2}\tmV^m}{m!}\mc{B}^{2*}_{(m)},\>\>\>
			\mc{K}_n(\ve^{1/2}\tmV) = \sum_{m=0}^\infty \frac{\ve^{m/2}\tmV^m}{m!}\mc{K}^{*}_{n;(m)}
		\end{equation}
		with $\mc{K}^*_{1;(0)}=0$. 
		In the limit $\ve \to 0$, we obtain the Fokker-Planck equation:
		\begin{equation}
			\frac{\partial \mc{P}(\tl{\mc{V}},t)}{\partial \tau} 	= \left[\tl{\gamma} \frac{\partial }{\partial \tmV}\tmV + \frac{\sigma^2 }{2}\frac{\partial^2 }{\partial \tmV^2}\right]\mc{P}(\tl{\mc{V}},t)  
																	+ \left[\gamma'\frac{\partial }{\partial \tmV}\tmV + \frac{\sigma'^2}{2}\frac{\partial^2 }{\partial \tmV^2}\right]\mc{P}(\tl{\mc{V}},t) ,\label{eq:FP_Gaussian}
		\end{equation}
		where $\tl{\gamma} \equiv A_{(1)}^*$, $\gamma' \equiv \mc{K}_{1;(1)}^*$, $\sigma^2 \equiv \mc{B}_{(0)}^{2*}$, and $\sigma'^2 \equiv \mc{K}_{2;(0)}^*$. 
		The Fokker-Planck equation~{(\ref{eq:FP_Gaussian})} is equivalent to the Gaussian Langevin equation as 
		\begin{equation}
			\frac{d\tmV}{d\tau} = -\tl{\gamma}\tmV + \sigma \hat \xi_G -\gamma' \tmV +\sigma' \hat \xi_G',\label{eq:Langevin_G}
		\end{equation}
		where $\hat \xi_G$ and $\hat \xi_G'$ are the independent white Gaussian noise terms satisfying $\la \hat \xi_G(\tau)\ra=\la \hat \xi'_G(\tau)\ra=0$ and $\la\hat \xi_G(\tau_1)\hat \xi_G(\tau_2)\ra=\la \hat \xi'_G(\tau_1)\hat \xi_G'(\tau_2)\ra=\delta(\tau_1-\tau_2)$. 
		Note that the frictional effect appears not only from the continuous force $\hat F_\ve^c$ but also from the discontinuous force $\hat F_\ve^d$ (see Fig.~\ref{fig:Assumption3}(b)). 
		In other words, the emergence of the Gaussian property is equivalent to the emergence of the frictional effect from the discontinuous force. 
		
	\subsection{Asymptotic connection from the non-Gaussian to the Gaussian theory}
		As we have shown, whether the system obeys the non-Gaussian Langevin equation~{(\ref{eq:NGL_NL_derived})} or the Gaussian one~{(\ref{eq:Langevin_G})} depends on the amplitude of the frictional effect in $\hat F^c_{\ve}(t)$.
		We here explain an asymptotic connection from the non-Gaussian Langevin equation~{(\ref{eq:NGL_NL_derived})} to the Gaussian one~{(\ref{eq:Langevin_G})} in terms of the amplitude of the frictional effect. 
		We first make the assumptions (i), (iv'), the linear friction $A_\ve (\hat v)=\gamma_{\ve} \hat v$, and the symmetric jump noise $\overline{W}(0;\mc{Y})=\overline{W}(0;-\mc{Y})$ (or equivalently, $K_{2n+1;(0)}^*=0$), 
		and restrict our analysis to the following two cases: 
		\begin{enumerate}
			\item {\it {The strong frictional case}}: $\gamma_\ve$ is positive and independent of $\ve$ , i.e., $\gamma_\ve=\gamma$. 
			In this case, the assumptions (ii) and (iii) are satisfied. 
			We then obtain
			\begin{equation}
				\frac{d\hat v}{dt} = -\gamma \hat v + \ve\sigma \hat \xi_{\rm G} + \ve \hat \xi_{\rm NG},\label{eq:NGL_connected_transient1}
			\end{equation}
			where $\hat \xi_G$ is the white Gaussian noise, 
			$\hat \xi_{\rm NG}$ is the white non-Gaussian noise characterized by the transition rate $\mc{W}(\mc{Y})=\overline{W}^*_{(0)}(\mc{Y})$, 
			and $\sigma^2\equiv \mc{B}_{(0)}^{2*}$. 
			We here use the original variable $\hat v$ as the representation. 
			
			\item {\it {The weak frictional case}}: $\gamma_\ve$ is scaled as $\gamma_\ve=\ve \tl{\gamma}$ with a positive and $\ve$-independent constant $\tl{\gamma}$. 
			In this case,  the assumptions (ii') and (iii') are satisfied. 
			We therefore obtain 
			\begin{equation}
				\frac{d\hat v}{dt} = -\ve\tl{\gamma} \hat v-\ve \gamma' \hat v + \ve \sigma \hat \xi_{\rm G} + \ve \sigma' \hat \xi'_{\rm G},\label{eq:NGL_connected_transient2}
			\end{equation}
			where $\gamma' = \mc{K}_{1;(1)}^*$, $\sigma'^2 = \mc{K}_{2;(0)}^*$, 
			and $\hat \xi_G$ and $\hat \xi_G'$ are the independent white Gaussian noise terms satisfying 
			$\la \hat \xi_G\ra=\la \hat \xi'_G\ra=0$ and $\la\hat \xi_G(t_1)\hat \xi_G(t_2)\ra=\la \hat \xi'_G(t_1)\hat \xi_G'(t_2)\ra=\delta(t_1-t_2)$.
			Note that we use the original variable $\hat v$ again as the representation.
		\end{enumerate}
		We note that the models~{(\ref{eq:NGL_connected_transient1})} and~{(\ref{eq:NGL_connected_transient2})} are not uniformly valid for the amplitude of $\gamma$. 
		
		We now propose the following single equation which is valid for both cases 1 and 2: 
		\begin{equation}
			\frac{d\hat v}{dt} = -\gamma_\ve \hat v -\ve \gamma' \hat v + \ve\sigma \hat \xi_{\rm G} + \ve \hat \xi_{\rm NG}.\label{eq:NGL_connected}
		\end{equation}
		In fact, Eq.~{(\ref{eq:NGL_connected})} is reduced to Eqs.~{(\ref{eq:NGL_connected_transient1})} and~{(\ref{eq:NGL_connected_transient2})} to leading order in terms of $\ve$ for the cases 1 and 2, respectively. 
		In the case 1, the second term on the rhs of Eq.~{(\ref{eq:NGL_connected})} is negligible because the typical value of $\hat v$ is the order of $\ve$ as shown in Sec.~{\ref{sec:NGL_nonliner}}, 
		which implies that Eq.~{(\ref{eq:NGL_connected})} is reduced to Eq.~{(\ref{eq:NGL_connected_transient1})} to leading order. 
		In the case 2, Eq.~{(\ref{eq:NGL_connected})} is reduced to Eq.~{(\ref{eq:NGL_connected_transient2})} as follows. 
		The Kramers-Moyal equation for Eq.~{(\ref{eq:NGL_connected})} is given by
		\begin{align}
			\frac{\partial P(v,t)}{\partial t} 	= \left[\ve \tl{\gamma}\frac{\partial }{\partial v}v + \frac{\ve^2\sigma^2}{2}\frac{\partial^2}{\partial v^2} + \ve \gamma'\frac{\partial }{\partial v}v + \sum_{n=1}^\infty \frac{\ve^{2n}\mc{K}_{2n;(0)}^*}{(2n)!}\frac{\partial^{2n}}{\partial v^{2n}}\right]P(v,t).
		\end{align}
		By introducing scaled variables $\tmV \equiv v/\sqrt{\ve}$, $\tau = \ve t$, and $\mc{P}(\tmV,\tau) \equiv \sqrt{\ve}P(v,t)$, 
		we obtain 
		\begin{align}
			\frac{\partial \mc{P}(\tmV,\tau)}{\partial \tau} 	= \left[\tl{\gamma}\frac{\partial }{\partial \tmV}\tmV + \frac{\sigma^2}{2}\frac{\partial^2}{\partial \tmV^2} + \gamma'\frac{\partial }{\partial \tmV}\tmV + \frac{\sigma'^2}{2}\frac{\partial^2}{\partial \tmV^2}\right]\mc{P}(\tmV,\tau) + \sum_{n=2}^\infty \frac{\ve^{n-1}\mc{K}_{2n;(0)}^*}{(2n)!}\frac{\partial^{2n}}{\partial \tmV^{2n}}\mc{P}(\tmV,\tau),
		\end{align}
		which implies Eq.~{(\ref{eq:FP_Gaussian})} in the limit $\ve \to 0$. 
		Equation~{(\ref{eq:NGL_connected})} is then equivalent to Eq.~{(\ref{eq:NGL_connected_transient2})} at leading order. 
		
	\subsection{Discussion on the validity of the non-Gaussian Langevin equation}
		We first remark the relationship of our formulation to the central limit theorem (CLT) and the non-equilibrium steady state. 
		We next generalize the concept of the non-linear temperature, which has been introduced in Ref.~\cite{Kanazawa1}, to show the explicit criteria where the small noise expansion is valid. 
		We also show that the small noise expansion fails to reproduce the tail of the distribution. 
		
		\subsubsection{Relation to the central limit theorem}
			We explain the relation between the CLT and our theory. 
			According to the CLT, the summation of independent and identically distributed (i.i.d) random variables asymptotically obeys the Gaussian distribution if all the cumulants of the i.i.d variables are finite. 
			Because the white non-Gaussian noise belongs to the class of the i.i.d random variables, 
			the simple summation of the white non-Gaussian noise $\int_0^t ds\hat \xi_{\rm NG}(s)/\sqrt{t}$ asymptotically converges to the Gaussian noise for $t\gg \tau_{\rm CLT}^*$ 
			with the characteristic time scale\footnote{
			The CLT time-scale $\tau_{\rm CLT}^*$ can be estimated to be $\tau_{\rm CLT}^*\simeq K_4/K_2^2$ with the second and fourth order cumulants $K_2$ and $K_4$.} $\tau_{\rm CLT}^*$. 
			If the relaxation time scale $\tau_{\rm S}^*$ is sufficiently long (i.e., $\tau_{\rm S}^* \gg \tau_{\rm CLT}^*$), 
			the system can be regarded as unchanged during time of the order $\tau_{\rm CLT}^*$ and the CLT is valid. 
			We therefore obtain the Gaussian Langevin equation~{(\ref{eq:Langevin_G})} under the condition (ii), which is physically equivalent to $\tau_{\rm S}^* \gg \tau_{\rm CLT}^*$. 
			On the other hand, if the relaxation time scale is not long enough as $\tau_{S}^* \lesssim \tau_{\rm CLT}^*$, the CLT is no longer applicable because the system changes its state during time of the order $\tau_{\rm CLT}^*$. 
			We then obtain the non-Gaussian Langevin equation~{(\ref{eq:NGL_NL_derived})} under the conditions (ii), which is physically equivalent to $\tau_{S}^* \lesssim \tau_{\rm CLT}^*$. 
		
		\subsubsection{Relation to the non-equilibrium steady state}
			The non-Gaussian Langevin equation~{(\ref{eq:NGL_NL_derived})} describes a system far from equilibrium because the local detailed balance condition is not satisfied. 
			To clarify this point, let us analyze the energy flux from the non-Gaussian to Gaussian bath on the basis of stochastic energetics~\cite{Sekimoto1,Sekimoto2,SekimotoB,Kanazawa2}. 
			The heat absorbed by the Gaussian bath is defined by $d\hat Q/dt \equiv (\gamma \hat{\mc{V}}-\sigma \hat \xi_{\rm G})\circ \hat{\mc{V}}$ with the Stratonovich product $\circ$~\cite{Gardiner}. 
			The heat flux then flows from the non-Gaussian to the Gaussian bath: $\la d\hat Q/dt\ra = K_2>0$, where $K_2$ is the second cumulant of the non-Gaussian noise $\hat\xi_{\rm NG}$. 
			Remarkably, the direction of heat flux is independent of $\sigma^2$ (i.e., the thermal temperature). 
			This result implies that the effective temperature of the non-Gaussian bath is much higher than that of the Gaussian bath. 
			Indeed, high temperature difference is shown necessary between the two baths in the example of a granular rotor associated with rarefied molecular gas (see Appendix~\ref{app:gmotor_viscous}). 
			We note that this condition is valid for systems where the non-Gaussian athermal fluctuations appear. 
			
		\subsubsection{Non-linear temperature}
			We here discuss the explicit criteria of the small noise assumption~{(i)} by introducing the concept of the non-linear temperature. 
			For simplicity, we make the assumptions (i) and (iv'), and consider the linear friction case $A_\ve(\hv)=\gamma \hv$ with an $\ve$-independent positive parameter~$\gamma$.
			We then expand $\mc{B}(\ve \mc{V})$ and $\overline{W}(\ve\mc{V};\mc{Y})$ as 
			\begin{equation}
				\mc{B}^2(\ve \mc{V})			= \sum_{n=0}^\infty \frac{\ve^{n}}{n!}\mc{V}^{n}\mc{B}_{(n)}^{2*},\>\>\>
				\overline{W}(\ve\mc{V};\mc{Y})	= \sum_{n=0}^\infty \frac{\ve^{n}}{n!}\mc{V}^{n}\overline{W}_{(n)}^*(\mc{Y}),\label{eq:der_exp2}
			\end{equation}
			where $\mc{B}^{2*}_{(1)}$ and $\overline{W}_{(1)}^*(\mc{Y})$ are assumed to be non-zero. 
			The essence of our expansion is to ignore the sub-leading terms as 
			\begin{align}
				|\mc{B}^{2*}_{(0)}| \gg \ve|\mc{B}^{2*}_{(1)}\mc{V}^{*}|, \>\>\> |\overline{W}_{(0)}^*(\mc{Y}^*)| \gg \ve^2|\overline{W}_{(1)}^*(\mc{Y}^*)\mc{V}^{*}|,\label{eq:smallnoise_criteria}
			\end{align}
			where $\mc{V}^*$ and $\mc{Y}^*$ are the typical values of $\mc{V}$ and $\mc{Y}$, respectively. 
			Note that the typical value of $\mc{V}$ relates to the effective temperature $\mc{T}$ as 
			\begin{equation}
				\mc{T} \equiv \frac{1}{2}\mc{V}^{*2} = \frac{\sigma^2 + \sigma'^2}{2\gamma},
			\end{equation}
			where $\sigma^2\equiv \mc{B}^{2*}_{(0)}$, $\sigma'^2\equiv \mc{K}_{2;(0)}^*$.
			Then, the condition~{(\ref{eq:smallnoise_criteria})} is equivalent to the low temperature condition: 
			\begin{equation}
				\mc{T}_{\rm NL} \gg \mc{T},
			\end{equation}
			where we have introduced the non-linear temperature 
			\begin{equation}
				\mc{T}_{\rm NL} \equiv \frac{1}{2\ve^2}\min\left(\left|\frac{\mc{B}^{2*}_{(0)}}{\mc{B}^{2*}_{(1)}}\right|, \left|\frac{\overline{W}_{(0)}^*(\mc{Y}^*)}{\overline{W}_{(1)}^*(\mc{Y}^*)}\right|^2\right). \label{eq:NL_temp}
			\end{equation}
			Note that the minimum function is defined as $\min(A,B)=A$ for $A\leq B$ and $\min(A,B)=B$ for $B<A$.
			The non-linear temperature~{(\ref{eq:NL_temp})} characterizes the temperature over which the non-linear terms in Eq.~{(\ref{eq:der_exp2})} become relevant. 
			
		\subsubsection{Tail of the distribution}
			We note that the Langevin-like description~{(\ref{eq:NGL_NL_derived})} is only valid for typical states of the system 
			(i.e., $|\mc{V}|\lesssim \mc{V}^*\equiv \sqrt{(\sigma^2+\sigma'^2)/\gamma}$) 
			and is invalid for rare states (i.e., $|\mc{V}|\gg \mc{V}^*$). 
			This is because the small noise expansion is not a uniform asymptotic expansion in terms of the velocity $\mc{V}$. 
			Indeed, for rare states $|\mc{V}|\gtrsim \mc{V}^*/\ve$, the higher-order terms in Eq.~{(\ref{eq:SME_NL_tran})} are not negligible anymore. 
			Fortunately, the probability of such rare trajectories is estimated to be extremely small, which ensures the validity of the Langevin-like description for typical trajectories. 
			We note that the same limitation also exists for the original theory of van Kampen (i.e., the Gaussian Langevin equation is also an effective description for typical trajectories).

\section{Asymptotic solutions for non-Gaussian Langevin equation with general non-linear friction}
	We have studied the derivation of the non-Gaussian Langevin equation~{(\ref{eq:NGL_NL_derived})}. 
	We next study their analytical solutions for the steady distribution function. 
	Because the exact solution for the linear case (i.e., $F(\hmV) = \gamma \hmV$) has been already obtained in Refs.~\cite{Eliazar,Kanazawa1}, 
	we study the non-linear frictional case and derive a full-order asymptotic formula in terms of the frictional coefficient. 
	We also show that the first-order truncation of the formula leads to the independent-kick model, which was phenomenologically introduced in Ref.~\cite{Talbot}. 
	We verify in detail the validity of the first-order formula for some specific cases: Coulombic and cubic frictions.  
	Furthermore, we introduce a diagrammatic representation for the multiple-kicks process during relaxation.

	\subsection{Setup}
		Let us consider the non-Gaussian Langevin equation with the non-linear friction~{(\ref{eq:NGL_NL_derived})}. 
		For simplicity, we focus on the case without the Gaussian noise $\sigma^2=0$. 
		We assume that the velocity $\mc{V}$ and time $t$ are nondimensionalized by the characteristic velocity of the friction function\footnote{
		For example, in the case with the cubic friction $f(\mc{V})=a\mc{V}+b\mc{V}^3$, the characteristic velocity scale of the friction function $f(\mc{V})$ is given by $\mc{V}^*\equiv \sqrt{a/b}$. } and the characteristic interval of the Poisson noises, respectively. 
		The steady distribution $\mc{P}_{\rm SS}(\mc{V})\equiv \lim_{t\to\infty}\mc{P}(\mc{V},t)$ satisfies 
		\begin{equation}
			\frac{\partial }{\partial \mc{V}}F(\mc{V})P_{\rm SS}(\mc{V}) + \int_{-\infty}^\infty d\mc{Y}\mc{W}(\mc{Y})\bigg\{\mc{P}_{\rm SS}(\mc{V-Y})-\mc{P}_{\rm SS}(\mc{V})\bigg\}=0.\label{eq:SME_NL_steady}
		\end{equation}
		We assume that Eq.~{(\ref{eq:SME_NL_steady})} has a unique solution satisfying $\mc{P}_{\SS}(\mc{V})\geq 0$ and $\int_{-\infty}^\infty d\mc{V}\mc{P}_{\rm SS}(\mc{V})=0$. 
		By introducing the Fourier representations 
		\begin{equation}
			\tl{P}(s) \equiv \int_{-\infty}^\infty d\mc{V} e^{is\mc{V}}\mc{P}_{\rm SS}(\mc{V}) \Longleftrightarrow \mc{P}_{\rm SS}(\mc{V}) \equiv \frac{1}{2\pi}\int_{-\infty}^\infty ds e^{-is\mc{V}}\tl{P}(s),
		\end{equation}
		and 
		\begin{equation}
			\tl{F}(s) \equiv \int_{-\infty}^\infty d\mc{V} e^{is\mc{V}}F(\mc{V}) \Longleftrightarrow F(\mc{V}) \equiv \frac{1}{2\pi}\int_{-\infty}^\infty ds e^{-is\mc{V}}\tl{F}(s).
		\end{equation}
		Equation~{(\ref{eq:SME_NL_steady})} is reduced to 
		\begin{equation}
			\frac{is}{2\pi}\int_{-\infty}^\infty du\tl{F}(s-u)\tl{P}(u) = \Phi(s)\tl{P}(s),\label{eq:IE_NL}
		\end{equation}
		where we have introduced the cumulant function
		\begin{equation}
			\Phi(s) \equiv \int_{-\infty}^\infty d\mc{Y}\mc{W}(\mc{Y})(e^{is\mc{Y}}-1).
		\end{equation}
		Our goal is to obtain the analytic solution of the linear integral equation~{(\ref{eq:IE_NL})}.

	\subsection{Asymptotic solution for strong friction}
\begin{figure}
			\centering
			\includegraphics[width=100mm]{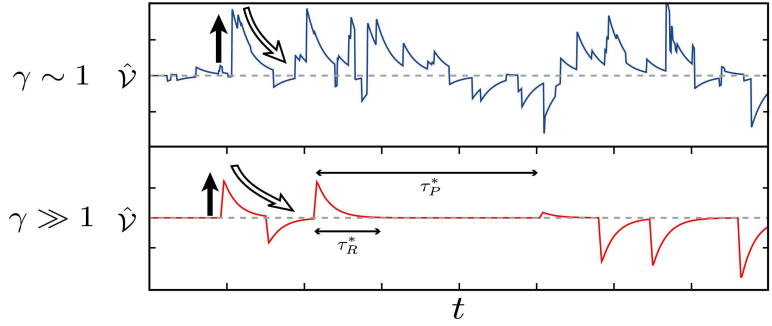}
			\caption{
						(Color online)
						Typical trajectories of the solution of the non-Gaussian Langevin equation~{(\ref{eq:NGL_NL_derived})} without the Gaussian noise ($\sigma^2=0$) for $\gamma \sim 1$ and $\gamma \gg 1$. 
						The filled and open arrows represent the excitation and relaxation due to the noise and the friction, respectively. 
						For $\gamma \gg 1$, the relaxation time scale $\tau_R^*$ is much shorter than the typical interval of the Poisson noise $\tau_P^*$, which implies that the system is localized at the rest state $\hat{\mc{V}}=0$. 
					}
			\label{fig:NG_trajectory}
		\end{figure}

		We here study the asymptotic expansion in terms of the inverse of the frictional coefficient.
		Let us assume that the friction function $F(\mc{V})$ is scaled by a positive large parameter $\gamma$ as 
		\begin{equation}
			F(\mc{V})=\gamma f(\mc{V}),
		\end{equation}
		where a typical trajectory of the tracer is illustrated in Fig.~{\ref{fig:NG_trajectory}}. 
		We note that the relaxation time scale $\tau_R^*$ is proportional to $\gamma^{-1}$, which implies that 
		$\gamma \gg 1$ is physically equivalent to $\tau_R^*/\tau_P^* \ll 1$ with the characteristic time interval of the Poisson noise $\tau_P^*$ (see Fig.~\ref{fig:NG_trajectory}). 
		We also assume that all integrals appropriately converge in the following calculations. 
		In the limit $\gamma\to\infty$, the steady distribution converges to the $\delta$-function around the stable point $\mc{V}=0$,
		i.e., $\lim_{\gamma\to\infty}\mc{P}_{\rm SS}(\mc{V})=\delta (\mc{V})$, which is equivalent to $\lim_{\gamma\to\infty}\tl{P}(s)=1$. 
		We then expand the Fourier representation $\tl{P}(s)$ in terms of the inverse of the friction coefficient $\mu\equiv 1/\gamma$ as 
		\begin{equation}
			\tl{P}(s) = 1 + \sum_{n=1}^\infty \mu^n \tl{a}_n(s),\label{eq:asym_postulate}
		\end{equation}
		where $\tl{a}_n(s)$ is a smooth function.  
		We note that $\tl{a}_n(s)$ satisfies the following relation because of the conservation of the probability:
		\begin{equation}
			\int_{-\infty}^\infty d\mc{V}\mc{P}_{\rm SS}(\mc{V})=\tl{P}(s=0)=1 \Longleftrightarrow \tl{a}_n(0)=0. 
		\end{equation}
		By introducing $\tl{f}(s)\equiv \int_{-\infty}^{\infty}d\mc{V} e^{is\mc{V}}f(\mc{V})$ 
		and substituting Eq.~{(\ref{eq:asym_postulate})} into Eq.~{(\ref{eq:IE_NL})}, 
		we obtain
		\begin{equation}
			 \frac{1}{2\pi}\int_{-\infty}^{\infty} du \tl{f}(s-u)\left[1 + \sum_{n=1}^\infty\mu^n \tl{a}_n(u)\right]
			= \frac{\mu\Phi(s)}{is}\left[1+\sum_{n=1}^\infty\mu^n \tl{a}_n(s)\right].
		\end{equation}
		From the assumption $\int_{-\infty}^\infty du \tl{f}(s-u)=f(0)=0$, we obtain an iterative relation for $\tl{a}_n$: 
		\begin{equation}
			\frac{1}{2\pi}\int_{-\infty}^{\infty} du\tl{f}(s-u)\tl{a}_{n+1}(u) = \frac{\Phi(s)}{is}\tl{a}_{n}(s),\label{eq:iterative_IE_trans}
		\end{equation}
		where we define $\tl{a}_0(s)=0$. 
		On the condition that $\tl{a}_n(0)=0$, Eq.~{(\ref{eq:iterative_IE_trans})} can be formally solved (see Appendix.~{\ref{app:CN:IntegralEq}} for details): 
		\begin{equation}
			\tl{a}_{n+1}(s) = \mc{I}[s;\tl{a}_n(s')],\label{eq:iterative_IE}
		\end{equation}
		where we have introduced a linear operator $\mc{I}$ for an arbitrary function $h(s')$ as
		\begin{equation}
			\mc{I}[s;h(s')] =\frac{1}{2\pi}\int_{-\infty}^{\infty}\frac{d\mc{V}(e^{is\mc{V}}-1)}{f(\mc{V})}\int_{-\infty}^{\infty}ds' e^{-is'\mc{V}}\frac{\Phi(s')}{is'}h(s'). 
		\end{equation}
		We then obtain the full-order asymptotic solution in terms of $\mu$: 
		\begin{equation}
			\tl{P}(s) 	= 1 + \mu \mc{I}[s;{\bm 1}(s')] + \mu^2 \mc{I}^2[s;{\bm 1}(s')]+\dots= [1-\mu\mc{I}]^{-1}[s;{\bm 1}(s')],\label{eq:Solution_Langevin_NLNG}
		\end{equation}
		where we introduce ${\bm 1}(s') = 1$ as the indicator function for the whole real-number space $\bm R^1$. 
		This formula is applicable to perturbatively calculate the steady distribution function for an arbitrary frictional force. 
		Furthermore, all of the terms in Eq.~{(\ref{eq:Solution_Langevin_NLNG})} can be physically interpreted as will be shown from the next subsections.
		
	\subsection{First-order approximation: the independent-kick model}
\begin{figure}
			\centering
			\includegraphics[height=25mm]{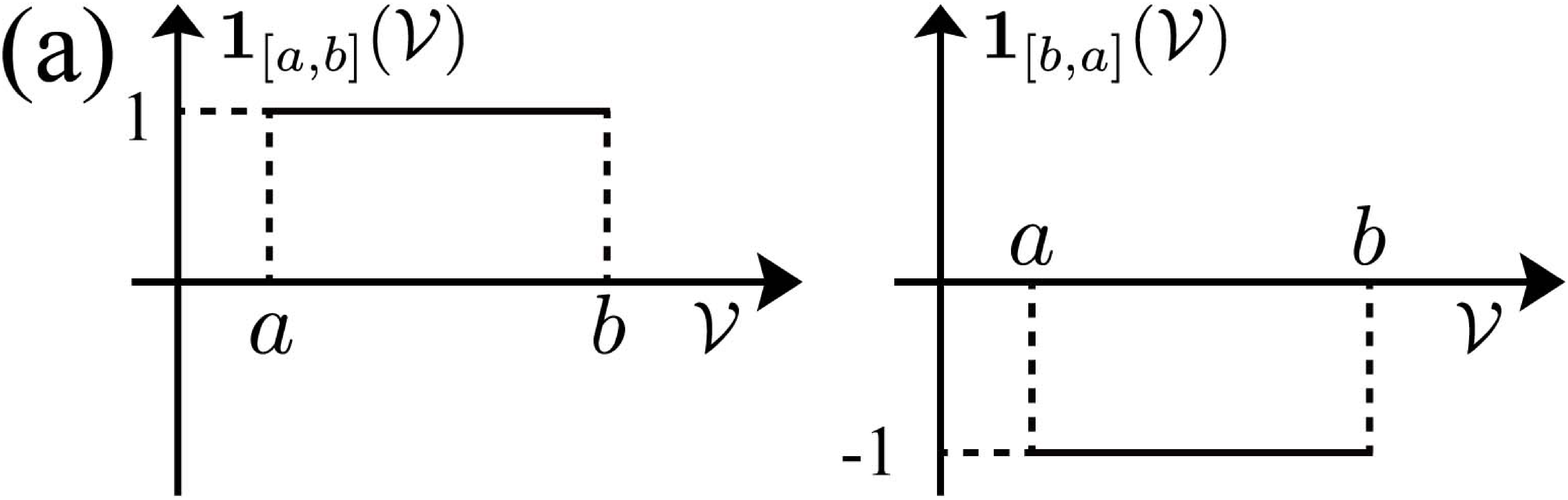}$\>\>\>\>$
			\includegraphics[height=25mm]{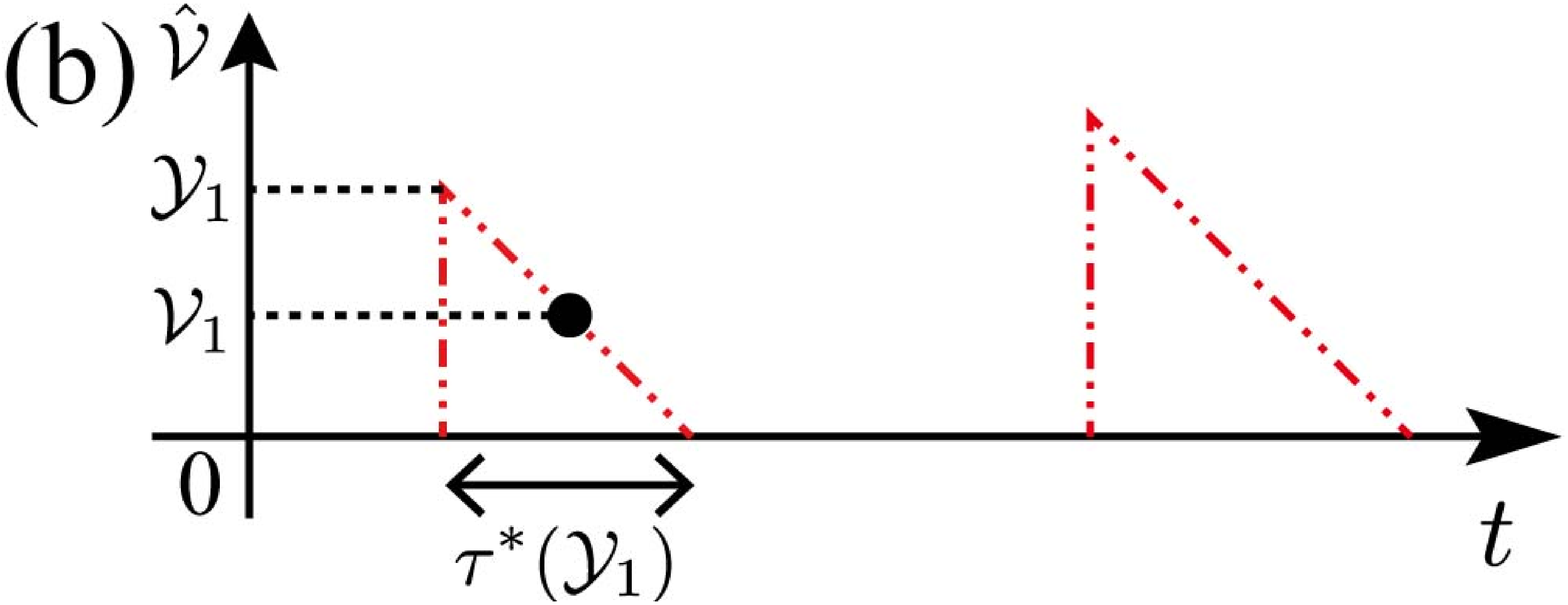}
			\caption{
						(Color online)
						(a)~Schematics of the indicator function ${\bm 1}_{[a,b]}(\mc{V})$ for $b\geq a$, where the indicator function takes the values $0$ or $\pm 1$. 
						(b)~Schematic of the independent-kick model (in the case of Coulombic friction $f(\mc{V})=\sgn(\mc{V})$). Because of the large friction, the system rapidly converges to the rest state ($\mc{V}$=0). The effect of multiple-kicks during relaxation is neglected. 
						The solid circle ($\bullet$) implies the integration in terms of $\mc{V}_1$. 
					}
			\label{fig:single_kicks}
		\end{figure}

		We here explain that the first-order asymptotic solution is equivalent to the independent-kick model, which was phenomenologically introduced in Ref.~\cite{Talbot}. 
		Let us first obtain the explicit form of the first-order solution as follows: 
		The explicit form of $\mc{I}[s;{\bm 1}(s')]$ can be simplified as
		\begin{align}
			\mc{I}[s;{\bm 1}(s')]
			&= \frac{1}{2\pi}\int_{-\infty}^{\infty}\frac{d\mc{V}(e^{is\mc{V}}-1)}{f(\mc{V})}\int_{-\infty}^{\infty}ds' \frac{ e^{-is'\mc{V}}\Phi(s')}{is'}\notag\\
			&= \frac{1}{2\pi}\int_{-\infty}^{\infty}\frac{d\mc{V}(e^{is\mc{V}}-1)}{f(\mc{V})}\int_{-\infty}^{\infty}ds' \frac{ e^{-is'\mc{V}}}{is'}\int_{-\infty}^{\infty}d\mc{Y} \mc{W}(\mc{Y})(e^{is'\mc{Y}}-1)\notag\\
			&= \int_{-\infty}^{\infty}d\mc{Y} \mc{W}(\mc{Y})\int_{-\infty}^{\infty}\frac{d\mc{V}(e^{is\mc{V}}-1)}{f(\mc{V})}\int_{-\infty}^{\infty}\frac{ds'}{2\pi} \frac{ e^{-is'\mc{V}}}{is'}(e^{is'\mc{Y}}-1)\notag\\
			&= \int_{-\infty}^{\infty}d\mc{Y} \mc{W}(\mc{Y})\int_{-\infty}^{\infty}\frac{d\mc{V}(e^{is\mc{V}}-1)}{f(\mc{V})}{\bm 1}_{[0,\mc{Y}]}(\mc{V})\notag\\
			&= \int_{-\infty}^{\infty}d\mc{Y}\mc{W}(\mc{Y})\int_{0}^{\mc{Y}}d\mc{V} \frac{e^{is\mc{V}}-1}{f(\mc{V})},
		\end{align}
		where we have introduced the indicator function (see Fig.~{\ref{fig:single_kicks}(a)}):
		\begin{equation}
			{\bm 1}_{[a,b]}(\mc{V})\equiv \frac{1}{2}[{\rm sgn}(\mc{V}-b)-{\rm sgn}(\mc{V}-a)].
		\end{equation}
		The indicator function takes the following values for $b\geq a$ as 
		\begin{equation}
			{\bm 1}_{[a,b]}(\mc{V}) = 	\begin{cases}
											0 & (x < a {\rm \>\>or\>\>} b<x) \\
											1 & (a \leq x \leq b)
										\end{cases}.
		\end{equation}
		We also note that the indicator function satisfies the relations for arbitrary numbers $a$ and $b$: 
		\begin{equation}
			{\bm 1}_{[a,b]}(\mc{V}) = -{\bm 1}_{[b,a]}(\mc{V}), \>\>\> \int_a^b dx f(x) = \int_{-\infty}^{\infty} dx f(x){\bm 1}_{[a,b]}(x).
		\end{equation}
		We then obtain the first-order asymptotic solution as 
		\begin{equation}
			\tl{P}(s) = 1 + \mu\int_{-\infty}^{\infty}d\mc{Y}\mc{W}(\mc{Y})\int_{0}^{\mc{Y}} \frac{d\mc{V}}{f(\mc{V})}[e^{is\mc{V}}-1] + O(\mu^2).\label{eq:first_approximation_independent_kick}
		\end{equation}
		
		We next show the first-order solution~{(\ref{eq:first_approximation_independent_kick})} can be interpreted as the independent-kick model~\cite{Talbot}.
		According to the physical picture of the independent-kick model, the system is typically in the rest state ($\mc{V}$=0) due to the large friction, but is sometimes excited by single-kicks. 
		We here assume that the system is not kicked during relaxation\footnote{This assumption is valid for the first-order approximation. 
		Modification due to higher-order corrections is discussed in Sec.~\ref{sec:multiplekicks}}, and is kicked only in the rest state (see Fig.~{\ref{fig:single_kicks}(b)} as a schematic of the independent-kick model).
		This scenario leads to the following formula: 
		\begin{equation}
			\la h(\hat{\mc{V}})\ra_{\rm SS} \simeq \int_{-\infty}^\infty d\mc{Y}\mc{W}(\mc{Y})\int_0^{\tau^*(\mc{Y})} dth(\mc{V}(t;\mc{Y})), \label{eq:independent_kick_general}
		\end{equation}
		where $h(\mc{V})$ is an arbitrary function of $\mc{V}$. 
		Here, $\mc{V}(t;\mc{Y})$ is the solution of the following differential equation on the initial condition $\mc{V}(0;\mc{Y})=\mc{Y}$ as
		\begin{equation}
			\frac{d\mc{V}}{dt} = -\frac{f(\mc{V})}{\mu} \Longleftrightarrow dt = -\mu\frac{d\mc{V}}{f(\mc{V})},\label{eq:diff_IK_trans}
		\end{equation}
		and $\tau^*(\mc{Y})$ is the stopping time defined by 
		\begin{equation}
			\tau^*(\mc{Y}) \equiv \mu\int_0^{\mc{Y}}\frac{d\mc{V}}{f(\mc{V})}.\label{eq:def_stopping_time}
		\end{equation}  
		We note that $\tau^*(\mc{Y})$ satisfies $\mc{V}(\tau^*;\mc{Y})=0$.
		We also note that $\tau^*(\mc{Y})$ can diverge (e.g., the case of the viscous friction $f(\mc{V})=\mc{V}$).
		By substituting $h(v)=e^{isv}-1$ into Eq.~{(\ref{eq:independent_kick_general})} and using Eq.~{(\ref{eq:diff_IK_trans})}, we obtain
		\begin{align}
			\tl{P}(s)-1 &\simeq \mu\int_{-\infty}^\infty d\mc{Y}\mc{W}(\mc{Y})\int_0^{\mc{Y}} \frac{d\mc{V}}{f(\mc{V})}\left[e^{is\mc{V}}-1\right],
		\end{align}
		which is equivalent to Eq.~{(\ref{eq:first_approximation_independent_kick})}. 
		We note that the assumption (iii) is implicitly used in the above calculation: 
		Under the assumption (iii), the differential equation~{(\ref{eq:diff_IK_trans})} has a monotonically decreasing (increasing) solution in terms of $t$ satisfying $\mc{V}(0;\mc{Y})=\mc{Y}$ and $\mc{V}(\tau^*(\mc{Y});\mc{Y})=0$ for a positive (negative) number $\mc{Y}$. 
		Therefore, the equation $\mc{V}(t;\mc{Y})=y$ can be implicitly solved uniquely for $t=t(y;\mc{Y})$.
		
	\subsection{Toy model 1: Coulombic friction}\label{subsec:toy1_Coulomb}
		We first consider the case with Coulombic friction and the symmetric jump force: 
		\begin{equation}
			f(\hat{\mc{V}}) = {\rm sgn} (\hat{\mc{V}}), \>\>\>\> \mc{W}(\mc{Y}) = \mc{W}(-\mc{Y}).\label{eq:Toymodel1}
		\end{equation}
		We note that, in this case, the tail of $\mc{P}_{\rm SS}(\mc{V})$ is given by the exponential form for an arbitrary $\mu$ as 
		\begin{equation}
			\mc{P}_{\rm SS}(\mc{V}) \sim e^{-a|\mc{V}|}\>\>\>\>(|\mc{V}|\gg \mc{D}^*),\label{eq:asymp_1st_coulomb_toy_dist}
		\end{equation}
		where $\mc{D}^*$ is the characteristic jump distance and $a$ is the solution of $a = \mu\Phi(-ia)$. 
		The asymptotic tail~{(\ref{eq:asymp_1st_coulomb_toy_dist})} can be shown as follows: 
		Assuming that the transition rate $\mc{W}(\mc{Y})$ decays sufficiently fast (i.e., $\mc{W}(\mc{Y})\to 0$ for $|\mc{Y}|\gg \mc{D}^*$), 
		we substitute Eq.~{(\ref{eq:asymp_1st_coulomb_toy_dist})} into the rhs of the master equation~{(\ref{eq:SME_NL_steady})} for $\mc{V} \gg \mc{D}^*$ to obtain
		\begin{align}
			&\frac{\partial }{\partial \mc{V}}\gamma {\rm sgn}(\mc{V})\mc{P}_{\rm SS}(\mc{V})
															+ \int_{-\infty}^\infty d\mc{Y}\mc{W}(\mc{Y})\bigg\{\mc{P}_{\rm SS}(\mc{V-Y})-\mc{P}_{\rm SS}(\mc{V})\bigg\}\notag\\
			\simeq& -\gamma ae^{-a\mc{V}} + \left[\int_{-\infty}^{\infty} d\mc{Y}\mc{W}(\mc{Y})(e^{a\mc{Y}}-1)\right]e^{-a\mc{V}}\notag\\
			=& [-\gamma a +\Phi(-ia)]e^{-a\mc{V}}=0.\label{eq:asymp_1st_coulomb_toy_dist_derivation}
		\end{align}
		Then, the asymptotic tail~{(\ref{eq:asymp_1st_coulomb_toy_dist})} satisfies the master equation~{(\ref{eq:SME_NL_steady})}. 
		
		We next study the first-order asymptotic solution in terms of $\mu$. 
		From Eq.~{(\ref{eq:first_approximation_independent_kick})}, the first-order asymptotic solution is given by
		\begin{equation}
			\tl{P}(s) = 1 + 2\mu \int_0^{\infty} d\mc{Y}\mc{W}(\mc{Y})\left[\frac{\sin{s\mc{Y}}}{s}-\mc{Y}\right] + O(\mu^2).\label{eq:1st_coulomb_toy}
		\end{equation}
		From Eq.~{(\ref{eq:1st_coulomb_toy})} and the relation $\la \hat{\mc{V}}^n\ra = (d^n\tl{P}(s)/d(is)^n)|_{s=0}$, 
		we obtain the moment as
		\begin{equation}
			\la \hat{\mc{V}}^{n}\ra = \begin{cases}
															\frac{2\mu}{n+1}\int_0^{\infty}d\mc{Y}\mc{W}(\mc{Y})\mc{Y}^{n+1} + O(\mu^2) 	& ({\rm for\>\> even\>\> }n) \\
															0																																		& ({\rm for\>\> odd\>\> }n)
														\end{cases}.
		\end{equation}
		
		Note that the asymptotic solution~{(\ref{eq:1st_coulomb_toy})} is uniformly valid in terms of $s$, 
		which implies that Eq.~{(\ref{eq:1st_coulomb_toy})} can be used to obtain the steady distribution function $\mc{P}_{\rm SS}(\mc{V})$. 
		Using the Fourier transformation, we obtain the steady distribution in the first-order approximation: 
		\begin{equation}				
			\mc{P}_{\rm SS}(\mc{V}) = \left[1-\mu c_1\right] \delta(\mc{V}) + \mu \int_0^\infty d\mc{Y}\mc{W}(\mc{Y}){\bm 1}_{[-\mc{Y},\mc{Y}]}(\mc{V})+O(\mu^2),\label{eq:1st_coulomb_toy_dist}
		\end{equation}
		where we have introduced $c_1\equiv 2\int_0^\infty d\mc{Y}\mc{W}(\mc{Y})\mc{Y}$. 
		The first term on rhs of Eq.~{(\ref{eq:1st_coulomb_toy_dist})} implies that the $\delta$-type singularity exists in the distribution function even for finite $\mu$,
		which emerges because Coulombic friction is sufficiently strong around $\mc{V}=0$ (i.e., the stopping time $\tau^*(\mc{Y})=\mu \mc{Y}$ is finite). 
		In the case with the symmetric Poisson noise $\mc{W}(\mc{Y}) = \lambda[\delta(\mc{Y}-\mc{Y}_0)+\delta(\mc{Y}+\mc{Y}_0)]/2$, we obtain 
		\begin{equation}
			\mc{P}_{\rm SS}(\mc{V}) = \left[1-\mu\lambda\mc{Y}_0\right] \delta(\mc{V}) + \frac{\mu\lambda}{2}{\bm 1}_{[-\mc{Y}_0,\mc{Y}_0]}(\mc{V}) + O(\mu^2). \label{eq:Coulomb_sPoisson}
		\end{equation}
		
		We here comment on the limitation of the first-order solutions~{(\ref{eq:1st_coulomb_toy_dist})} and~{(\ref{eq:Coulomb_sPoisson})}. 
		The asymptotic solution~{(\ref{eq:1st_coulomb_toy_dist})} is not uniformly valid for $\mc{V}$, 
		and is only valid for $|\mc{V}| \lesssim \mc{D}^*$ with the characteristic jump distance $\mc{D}^*$. 
		This is because the first-order solution~{(\ref{eq:1st_coulomb_toy_dist})} corresponds to the independent-kick picture. 
		If we are interested in the behavior for $|\mc{V}|\gg\mc{D}^*$, we have to take into account the multiple-kicks effect during relaxation, which will be studied in Sec.~{\ref{sec:multiplekicks}}. 
		Indeed, the tail form~{(\ref{eq:asymp_1st_coulomb_toy_dist})} is totally different from the independent-kick solution~{(\ref{eq:1st_coulomb_toy_dist})}.
		
		\subsubsection*{Numerical validation}
 		\begin{figure}
			\centering
			\includegraphics[width=150mm]{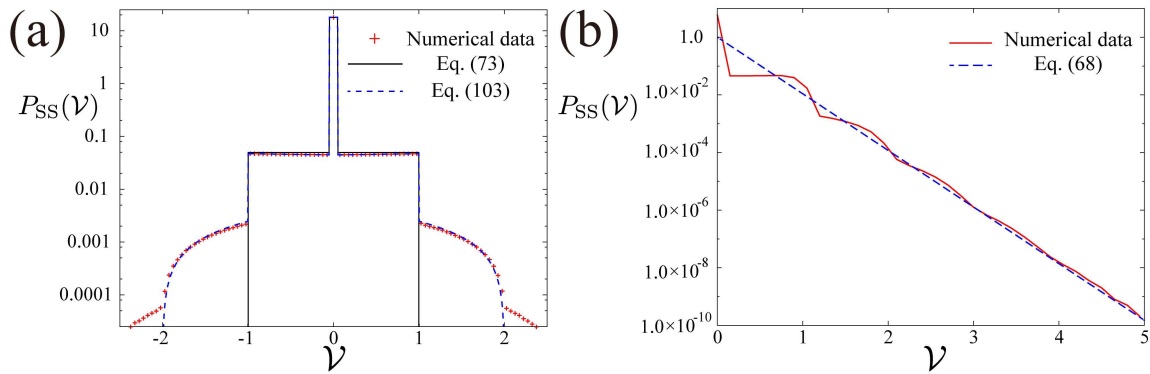}
			\caption{
						(Color online)
						Numerical validation of the formulas~{(\ref{eq:Coulomb_sPoisson})}, {(\ref{eq:asymp_1st_coulomb_toy_dist})}, and~{(\ref{eq:sol_Toymodel1_2nd})}. 
						(a) We compare the numerical data (red cross points), the first-order solution~{(\ref{eq:Coulomb_sPoisson})} (solid line), and the second-order solution~{(\ref{eq:sol_Toymodel1_2nd})} (dashed line). 
							The ensemble number of the Monte Carlo simulation is approximately $2.3\times 10^9$.
							The first-order and second-order solutions~{(\ref{eq:Coulomb_sPoisson})} and~{(\ref{eq:sol_Toymodel1_2nd})} are valid only in the ranges $|\mc{V}|\leq \mc{Y}_0=1$ and $|\mc{V}|\leq 2\mc{Y}_0=2$, respectively. 
						(b) We compare the numerical data (solid line) and the asymptotic tail~{(\ref{eq:asymp_1st_coulomb_toy_dist})} (dashed line). 
							The ensemble number of the Monte Carlo simulation is approximately $2.4\times 10^{11}$.
					}
			\label{fig:Coulomb_Poisson}
		\end{figure}
		We have numerically checked the validity of the formulas~{(\ref{eq:Coulomb_sPoisson})} and~{(\ref{eq:asymp_1st_coulomb_toy_dist})}.
		We perform the Monte Carlo simulation for the symmetric Poisson noise $\mc{W}(\mc{Y}) = \lambda[\delta(\mc{Y}-\mc{Y}_0)+\delta(\mc{Y}+\mc{Y}_0)]/2$ 
		with parameters $\gamma=10$, $\lambda=1$ and $\mc{Y}_0=1$. 
		In Fig.~{\ref{fig:Coulomb_Poisson}(a)}, we demonstrate that 
		the first-order solution~{(\ref{eq:Coulomb_sPoisson})} is only valid for $|\mc{V}| \leq \mc{Y}_0=1$ as shown in Fig.~{\ref{fig:Coulomb_Poisson}(a)}. 
		In Fig.~{\ref{fig:Coulomb_Poisson}(b)}, we also show that the asymptotic tail of the distribution function can be well-described by the exponential function~{(\ref{eq:asymp_1st_coulomb_toy_dist})}. 
		We note that the second-order solution~{(\ref{eq:sol_Toymodel1_2nd})}, which is illustrated in Fig.~{\ref{fig:Coulomb_Poisson}(a)}, will be discussed in detail in Sec.~\ref{sec:multiplekicks}.

	\subsection{Toy model 2: the cubic friction}\label{subsec:toy2_Cubic}
		Let us consider the case of the cubic friction and the symmetric noise:
		\begin{equation}
			f(\hat{\mc{V}}) = \hat{\mc{V}} + \hat{\mc{V}}^3, \>\>\>\> \mc{W}(\mc{Y}) = \mc{W}(-\mc{Y}).
		\end{equation}
		In this case, the integral equation~{(\ref{eq:IE_NL})} is reduced to the third-order ordinary differential equation: 
		\begin{equation}
			\left[\frac{d}{ds} - \frac{d^3}{ds^3}\right]\tl{P}(s) = \frac{\mu\Phi(s)}{s}\tl{P}(s).\label{eq:ODE_cubic_toy}
		\end{equation}
		The asymptotic tail of $\tl{P}(s)$ for an arbitrary $\mu$ is given by 
		\begin{equation}
			\tl{P}(s) \simeq C \exp{\left[2\mu\int_0^\infty d\mc{Y}\mc{W}(\mc{Y})\int_0^\mc{Y}d\mc{V} \frac{\cos{s\mc{V}}-1}{\mc{V}}\right]}\sim |s|^{-\mu\lambda^*}\to 0 \>\>\>\>\>\> (s\to\infty), \label{eq:asymptotic_tail_cubic}
		\end{equation}
		where $\lambda^*\equiv 2\int_0^{\infty}d\mc{Y}\mc{W}(\mc{Y})$ and $C$ is an appropriate constant (see Appendix~{\ref{app:CN:tail_cubic}} for detail). 
		We note that the cubic friction is sufficiently weak around $\mc{V}=0$ and the stopping time diverges to infinity as $\tau^*(\mc{Y})\to \infty$. 
		This implies that the velocity of the tracer cannot exactly stay rest at $\mc{V}=0$, but distributes around $\mc{V}=0$. 
		Then, the steady distribution function $\mc{P}_{\rm SS}(\mc{V})$ has no singular part or, equivalently, its Fourier representation $\tl{P}(s)$ belongs to the class of the $L^1$-functions. 
		Indeed, the asymptotic form~{(\ref{eq:asymptotic_tail_cubic})} implies that $\mc{P}_{\rm SS}(\mc{V})$ asymptotically diverges around the stable zero point $\mc{V}=0$ for $\mu\lambda^* < 1$ as  
		\begin{equation}
			\mc{P}_{\rm SS}(\mc{V}) \sim |\mc{V}|^{-(1-\mu\lambda^*)}, \label{eq:divergence_cubic}
		\end{equation}
		because $\tilde{P}(s)\sim |s|^{-\mu\lambda^*}$ for $|s|\gg 1$. 
		We also note that the asymptotic form~{(\ref{eq:asymptotic_tail_cubic})} implies the existence of a transition point $\mu\lambda^*=1$. 
		For $\mu\lambda^* > 1$, the distribution $\mc{P}_{\rm SS}(\mc{V})$ is regular around $\mc{V}=0$,
		while $\mc{P}_{\rm SS}(\mathcal{V})$ diverges at $\mathcal{V}=0$ for $\mu\lambda^*<1$.
		
		Here, let us calculate the first-order asymptotic solution in terms of $\mu$.
		From Eq.~{(\ref{eq:first_approximation_independent_kick})}, we obtain
		\begin{equation}
			\tl{P}(s) 	= 1 + 2\mu\int_0^\infty d\mc{Y}\mc{W}(\mc{Y})\int_0^\mc{Y}d\mc{V} \frac{\cos{s\mc{V}}-1}{\mc{V}(1+\mc{V}^2)} + O(\mu^2).\label{eq:asymp_1st_cubic_toy}
		\end{equation}
		From the formula $\la \hat{\mc{V}}^{n}\ra=(d^n\tl{P}(s)/d(is)^n)|_{s=0}$,
		we obtain the moment $\la \hat{\mc{V}}^{n}\ra$ as 
		\begin{equation}
			\la \hat{\mc{V}}^{n}\ra = 	\begin{cases}
																	(-1)^{n/2}\mu\int_0^\infty d\mc{Y}\mc{W}(\mc{Y})\beta_{-\mc{Y}^2}(n/2,0)+O(\mu^2) 		& ({\rm for\>\> even\>\>}n) \\
																	0																																				 		& ({\rm for\>\> odd\>\>}n)
																\end{cases},\label{eq:asymp_1st_cubic_toy_moments}
		\end{equation}
		where $\beta_{z}(a,b)\equiv \int_0^zdt t^{a-1}(1-t)^{b-1}$ is the incomplete beta function. 
		For the symmetric Poisson noise $\mc{W}(\mc{Y}) = \lambda[\delta(\mc{Y}-\mc{Y}_0)+\delta(\mc{Y}+\mc{Y}_0)]/2$, 
		Eq.~{(\ref{eq:asymp_1st_cubic_toy_moments})} is reduced to $\la \hat{\mc{V}}^{n}\ra = (-1)^{n/2}\mu\beta_{-\mc{Y}^2_0}(n/2,0)/2+O(\mu^2)$ for even $n$.
		
		Unfortunately, the asymptotic expansion~{(\ref{eq:asymp_1st_cubic_toy})} is not uniformly valid for $s$
		because the second term on the rhs of Eq.~{(\ref{eq:asymp_1st_cubic_toy})} diverges in the limit $s\rightarrow \pm\infty$: 
		\begin{align}
			&\int_0^\infty d\mc{Y}\mc{W}(\mc{Y})\int_0^\mc{Y}d\mc{V} \frac{\cos{s\mc{V}}-1}{\mc{V}(1+\mc{V}^2)}
			=\int_0^\infty d\mc{Y}\mc{W}(\mc{Y})\int_0^\mc{Y}d\mc{V} \left[\frac{\cos{s\mc{V}}-1}{\mc{V}} -  \frac{\mc{V}(\cos{s\mc{V}}-1)}{1+\mc{V}^2}\right]\notag\\
			&\simeq \int_0^\infty d\mc{Y}\mc{W}(\mc{Y})\left[ -{\rm Cin}(s\mc{Y}) + \int_0^\mc{Y}\frac{\mc{V}d\mc{V}}{1+\mc{V}^2}\right]
			= \int_0^\infty d\mc{Y}\mc{W}(\mc{Y})\left[ -{\rm Cin}(s\mc{Y}) + \frac{1}{2}\log{(1+\mc{Y}^2)}\right],
		\end{align}
		where ${\rm Cin}(x)\equiv \int_{0}^{x}dt (1-\cos{t})/t$ is the cosine integral and we have used the Riemann-Lebesgue lemma $\lim_{s\to \infty}\int_0^\mc{Y}d\mc{V} \mc{V}\cos{s\mc{V}}/(1+\mc{V}^2)=0$~\cite{RL_lemma}. 
		From the asymptotic form of the cosine integral as ${\rm Cin}(x) = \log{x} + O(1)$ for $x\gg 1$, 
		the second term on the rhs of Eq.~{(\ref{eq:asymp_1st_cubic_toy})} diverges in the limit $s\to \infty$ as $\int_0^\infty d\mc{Y}\mc{W}(\mc{Y})\int_0^\mc{Y}d\mc{V} (\cos{s\mc{V}}-1)/(\mc{V}(1+\mc{V}^2)) \sim -(\lambda^*/2) \log{s} \to-\infty$.
		In order to renormalize this secular term, we here assume the following first-order solution 
		\begin{equation}
			\tl{P}(s) = \exp \left[2\mu\int_0^\infty d\mc{Y}\mc{W}(\mc{Y})\int_0^\mc{Y}d\mc{V} \frac{\cos{s\mc{V}}-1}{\mc{V}(1+\mc{V}^2)}\right] + O(\mu^2).\label{eq:asymp_1st_cubic_toy_RN}
		\end{equation}
		Note that the renormalized solution~{(\ref{eq:asymp_1st_cubic_toy_RN})} reproduces the asymptotic tail~{(\ref{eq:asymptotic_tail_cubic})} for $s\to \infty$ as 
		\begin{equation}
					\exp \left[2\mu\int_0^\infty d\mc{Y}\mc{W}(\mc{Y})\int_0^\mc{Y}d\mc{V} \frac{\cos{s\mc{V}}-1}{\mc{V}(1+\mc{V}^2)}\right]
			\simeq		C'\exp \left[2\mu\int_0^\infty d\mc{Y}\mc{W}(\mc{Y})\int_0^\mc{Y}d\mc{V} \frac{\cos{s\mc{V}}-1}{\mc{V}}\right],
		\end{equation}
		where we have introduced a constant $C'\equiv \exp[\mu\int_0^\infty d\mc{Y}\mc{W}(\mc{Y})\log{(1+\mc{Y}^2)}]$. 
		Furthermore, Eq.~{(\ref{eq:asymp_1st_cubic_toy_RN})} satisfies the original differential equation~{(\ref{eq:ODE_cubic_toy})} without divergence even in the limit $s\to \infty$ as shown in Appendix~{\ref{app:CN:renormalized}}.
		From Eq.~{(\ref{eq:asymp_1st_cubic_toy_RN})}, we obtain the steady distribution function:
		\begin{equation}
			\mc{P}_{\rm SS}(\mc{V}) = \int_{-\infty}^\infty \frac{ds}{2\pi} \exp \left[-is\mc{V}+2\mu\int_0^\infty d\mc{Y}\mc{W}(\mc{Y})\int_0^\mc{Y}d\mc{V} \frac{\cos{s\mc{V}}-1}{\mc{V}(1+\mc{V}^2)}\right] + O(\mu^{2}).\label{eq:asymp_1st_cubic_toy_dist}
		\end{equation}
		We note that Eq.~{(\ref{eq:asymp_1st_cubic_toy_dist})} is only valid for $|\mc{V}| \lesssim \mc{D}^*$ with the characteristic jump distance $\mc{D}^*$ 
		because the first-order solution~{(\ref{eq:asymp_1st_cubic_toy_RN})} corresponds to the independent-kick model. 
		
		\subsubsection*{Numerical validation}
			\begin{figure}
				\centering
				\includegraphics[width=150mm]{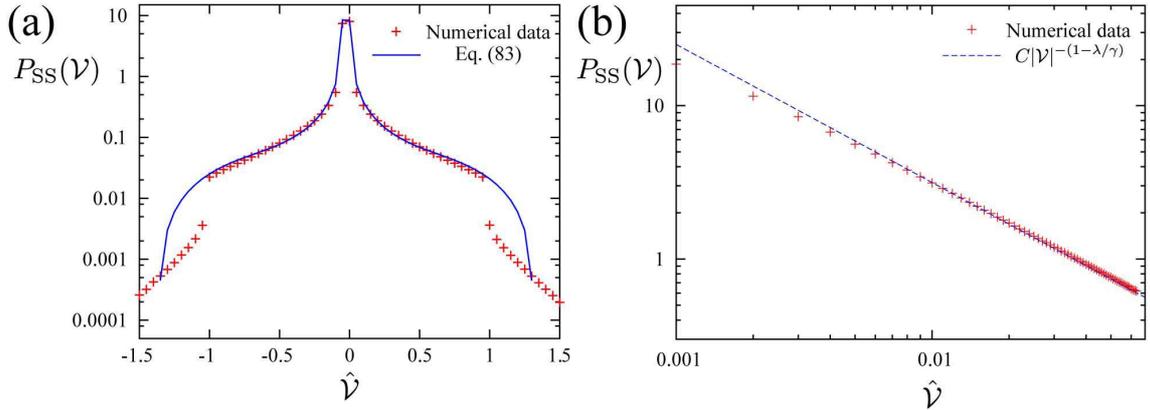}
				\caption{
							(Color online)
							(a) Comparison between Eq.~{(\ref{eq:asymp_1st_cubic_toy_dist})} and the numerical data of $\mc{P}_{\rm SS}(\mc{V})$ obtained by the Monte Carlo simulation for 
								the symmetric Poisson noise with parameters $\gamma=10$, $\lambda=1$, and $\mc{Y}_0=1$.
								The time step in the simulation and the ensemble number are set to be $10^{-3}$ and $2.18 \times 10^8$, respectively. 
								The renormalized solution~{(\ref{eq:asymp_1st_cubic_toy_dist})} is consistent with the numerical data in the range $|\mc{V}|\leq \mc{Y}_0=1$. 
							(b) The asymptotic form~{(\ref{eq:divergence_cubic_toy})} and the numerical data of $\mc{P}_{\rm SS}(\mc{V})$ around the peak 
								for the symmetric Poisson noise with parameters $\gamma=10$, $\lambda=1$, and $\mc{Y}_0=1$.
								The constant $C$, the time step in the simulation and the ensemble number are set to be $1/20$, $10^{-3}$, and $2.18 \times 10^8$, respectively. 
						}
				\label{fig:Cubic_Poisson}
			\end{figure}
			We numerically verify the validity of the first-order solution for the symmetric Poisson noise $\mc{W}(\mc{Y})=\lambda [\delta(\mc{Y}-\mc{Y}_0)+\delta(\mc{Y}+\mc{Y}_0)]/2$
			with parameters $\gamma=10$, $\lambda=1$, and $\mc{Y}_0=1$. 
			In this case, Eq.~{(\ref{eq:asymp_1st_cubic_toy_RN})} can be rewritten as 
			\begin{equation}
				\tl{P}(s) = \exp \left[\mu\lambda\int_0^{\mc{Y}_0}d\mc{V} \frac{\cos{s\mc{V}}-1}{\mc{V}(1+\mc{V}^2)}\right]=\exp \left[\mu\lambda\left(\frac{\log{(1+\mc{Y}_0^2)}}{2}-{\rm Cin}(s\mc{Y}) -\int_0^{\mc{Y}_0}\frac{\mc{V}\cos s\mc{V}d\mc{V}}{1+\mc{V}^2}\right)\right] + O(\mu^2).
			\end{equation}
			The asymptotic form of the peak $\mc{V}=0$ is given by 
			\begin{equation}
				\mc{P}_{\rm SS}(\mc{V}) \sim |\mc{V}|^{-(1-\mu\lambda)}, \label{eq:divergence_cubic_toy}
			\end{equation}
			for $-1< -(1-\mu\lambda) < 0$, where $\mc{P}_{\rm SS}(\mc{V})$ is integrable around the peak. 
			To verify the validity of Eqs.~{(\ref{eq:asymp_1st_cubic_toy_dist})} and~{(\ref{eq:divergence_cubic_toy})},	
			we perform the Monte Carlo simulation to obtain the numerical distribution function $\mc{P}_{\rm SS}(\mc{V})$ and compared it with Eqs.~{(\ref{eq:asymp_1st_cubic_toy_dist})} and~{(\ref{eq:divergence_cubic_toy})}.
			As shown in Fig.~{\ref{fig:Cubic_Poisson}}, our formulas~{(\ref{eq:asymp_1st_cubic_toy_dist})} and~{(\ref{eq:divergence_cubic_toy})} agree with our simulation for $|\mc{V}|\leq \mc{Y}_0$. 
			We also note that the power-law peak~{(\ref{eq:divergence_cubic_toy})} for the cubic friction is quite different from the $\delta$-type peak~{(\ref{eq:Coulomb_sPoisson})} for Coulombic friction. 
			This power-law originates from the divergence of the stopping time as $\tau^*(\mc{Y})\to\infty$. 

	\subsection{Higher-order corrections: multiple-kicks}\label{sec:multiplekicks}
		We here explicitly write the higher-order correction terms to illustrate their physical meaning. 
		We first note the following identity: 
		\begin{equation}
			\mc{I}[s;e^{i\mc{V}'s'}] = \int_{-\infty}^\infty d\mc{Y}\mc{W}(\mc{Y})\int_{-\infty}^{\infty}\frac{d\mc{V}}{f(\mc{V})}{\bm 1}_{[\mc{V'},\mc{V'+Y}]}(\mc{V})(e^{is\mc{V}}-1). \label{eq:relation_exp_multi_kick}
		\end{equation}
		Higher-order correction terms can be obtained using this identity.  
		For example, the second-order formula is given by
		\begin{equation}
			\mu^2\mc{I}^2[s;{\bm 1}(s')] = \int_{-\infty}^{\infty} d\mc{Y}_1\mc{W}(\mc{Y}_1)\frac{\mu d\mc{V}_1}{f(\mc{V}_1)}d\mc{Y}_2\mc{W}(\mc{Y}_2)\frac{\mu d\mc{V}_2}{f(\mc{V}_2)}(e^{is\mc{V}_2}-1)
										{\bm 1}_{[0,\mc{Y}_1]}(\mc{V}_1)[{\bm 1}_{[\mc{V}_1,\mc{V}_1+\mc{Y}_2]}(\mc{V}_2)-{\bm 1}_{[0,\mc{Y}_2]}(\mc{V}_2)].\label{eq:2nd-multi_kick}
		\end{equation}
		Introducing the abbreviation $d\Gamma_i\equiv \mu d\mc{Y}_i\mc{W}(\mc{Y}_i)d\mc{V}_i/f(\mc{V}_i)$ and ${\bm 1}_{[1,b]}^{(i)}\equiv{\bm 1}_{[1,b]}(\mc{V}_i)$,
		the explicit formula for the $n$th-order term with an integer $n\geq 2$ is represented as 
		\begin{equation}
			\mu^n\mc{I}^n[s;{\bm 1}(s')]=\int_{-\infty}^{\infty} d\Gamma_1{\bm 1}_{[0,\mc{Y}_1]}^{(1)}\prod_{i=2}^n \left[d\Gamma_i [{\bm 1}_{[\mc{V}_{i-1},\mc{V}_{i-1}+\mc{Y}_{i}]}^{(i)}-{\bm 1}_{[0,\mc{Y}_{i}]}^{(i)}]\right](e^{is\mc{V}_n}-1).\label{eq:nth-multi_kick}
		\end{equation}
	
		\begin{figure}
			\centering
			\includegraphics[width=150mm]{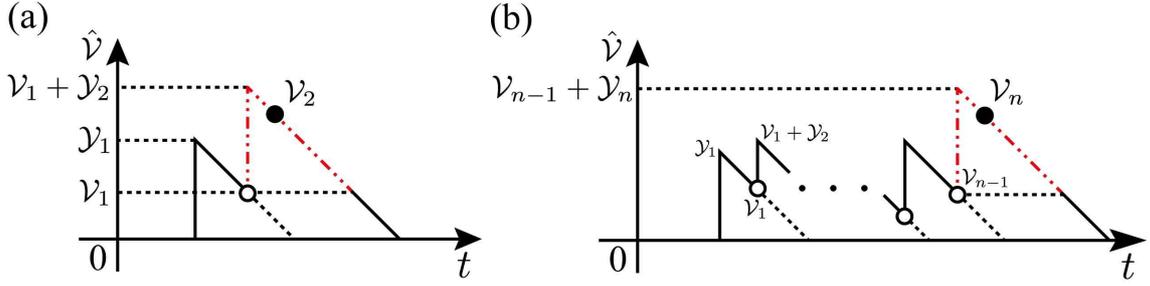}
			\caption	{
							(Color online)
							(a) A typical trajectory along which the system is kicked two times during relaxation. The two-dot-dash line corresponds to the diagram $(\circ \to \bullet)$.
							(b) A typical trajectory along which the system is kicked $n$ times during relaxation. The two-dot-dash line corresponds to the diagram $(\circ^{n-1} \to \bullet)$ 
							}
			\label{fig:multiple_kicks_model}
		\end{figure}
		We next discuss the physical meaning of the higher-order terms by introducing a diagram representation. 
		In the following, we restrict our theory to the case where $\tau^*(\mc{Y})$ is finite for an arbitrary finite number $\mc{Y}$ (e.g., Coulombic friction case). 
		For simplicity, we first consider the second-order formula~{(\ref{eq:2nd-multi_kick})}. 
		The second-order solution can be rewritten as
		\begin{align}
			\tl{P}(s) = 1 + [1- (\circ)](\bullet) + (\circ \to \bullet) +O(\mu^{3}),\label{eq:second_approximation_kick}
		\end{align}
		where we have introduced the following diagrams:
		\begin{align}
			(\bullet) 		&= \int_{-\infty}^\infty d\mc{Y}_1\mc{W}(\mc{Y}_1)\int_0^{\mc{Y}_1}\frac{\mu d\mc{V}_1}{f(\mc{V}_1)}(e^{is\mc{V}_1}-1) = \mc{I}_0[0;e^{is\mc{V}'}-{\bm 1}(\mc{V}')],\label{eq:diagram_ex1}\\
			(\circ) 			&= \int_{-\infty}^\infty d\mc{Y}_1\mc{W}(\mc{Y}_1)\int_0^{\mc{Y}_1}\frac{\mu d\mc{V}_1}{f(\mc{V}_1)} = \mc{I}_0[0;{\bm 1}(\mc{V}')],\label{eq:diagram_ex2}\\
			(\circ \to \bullet) &= \int_{-\infty}^\infty \!\!\!d\mc{Y}_1\mc{W}(\mc{Y}_1)\!\!\int_0^{\mc{Y}_1}\!\!\!\frac{\mu d\mc{V}_1}{f(\mc{V}_1)}\!\!\int_{-\infty}^\infty \!\!\!d\mc{Y}_2\mc{W}(\mc{Y}_2)\!\!\int_{\mc{V}_1}^{\mc{V}_1+\mc{Y}_2}\!\!\!\frac{\mu d\mc{V}_2}{f(\mc{V}_2)}(e^{is\mc{V}_2}\!-\!1)
								\!=\!\mc{I}_0^2[0;e^{is\mc{V}'}\!-\!{\bm 1}(\mc{V}')]. \label{eq:diagram_ex3}
		\end{align}
		We here also introduce the linear operator for an arbitrary function $h(\mc{V}')$ as 
		\begin{equation}
			\mc{I}_0[\mc{V};h(\mc{V}')] \equiv \int_{-\infty}^\infty d\mc{Y} \mc{W}(\mc{Y})\int_{\mc{V}}^{\mc{V}+\mc{Y}} \frac{\mu d\mc{V}'}{f(\mc{V}')}h(\mc{V}').
		\end{equation}
		The symbol $\bullet$ denotes the bound variable coupled with the exponential factor $(e^{is\mc{V}}-1)$ in the integrals (e.g., $\mc{V}_1$ in Eq.~{(\ref{eq:diagram_ex1})} and $\mc{V}_2$ in Eq.~{(\ref{eq:diagram_ex3})}), 
		the symbol $\circ$ denotes the bound variable decoupled of the exponential factor in the integrals (e.g., $\mc{V}_1$ in Eqs.~{(\ref{eq:diagram_ex2})} and {(\ref{eq:diagram_ex3})}), and  
		the arrow $\to$ represents that the limits of the latter integral is a function of the bound variable in the former integral
		(e.g., the integration range for $\mc{V}_2$ is designated by $\mc{V}_1$ as $\mc{V}_2\in [\mc{V}_1,\mc{V}_1+\mc{Y}_2]$ in Eq.~{(\ref{eq:diagram_ex3})}). 
		The diagram $(\bullet)$ corresponds to the effect of the single-kicks trajectories (the two-dot-dash line trajectory in Fig.~{\ref{fig:single_kicks}(b)}), where $\mc{V}_1$ moves along the two-dot-dash line. 
		Note that the diagram $(\bullet)$ depends on $s$ through the factor $(e^{is\mc{V}}-1)$. 
		From Eq.~{(\ref{eq:def_stopping_time})}, on the other hand, the diagram $(\circ)$ can be rewritten as (the integral along the solid line in Fig.~\ref{fig:multiple_kicks_model}(a)):
		\begin{equation}
			(\circ)=\int_{-\infty}^\infty d\mc{Y}_1\mc{W}(\mc{Y}_1)\tau^*(\mc{Y}_1),
		\end{equation}
		which is the probability that the second kick takes place during relaxation to leading order. 
		Note that the diagram $(\circ)$ is just a constant. 
		Furthermore, the diagram $(\circ \to \bullet)$ represents the effect of the second kick during relaxation (the two-dot-dash line in Fig.~{\ref{fig:multiple_kicks_model}(a)}),
		where $\mc{Y}_1$ is the initial condition from the first kick, $\mc{V}_1 \in [0,\mc{Y}_1]$ is the velocity before the second kick, and 
		$\mc{Y}_2$ is the flight distance by the second kick, and $\mc{V}_2\in [\mc{V}_1,\mc{V}_1+\mc{Y}_2]$ moves along the trajectory after the second kick. 
		Thus, the term $[1-(\circ)]$ represents the probability that the second kick does not occur during relaxation, and is used to modify the effect of the single-kicks trajectories up to the second-order as $[1-(\circ)](\bullet)$.  
		The term $(\circ \to \bullet)$ represents the direct contribution of the double-kicks trajectory. 
		
		\begin{table}
			\caption{Summary of the rules of the diagrams. The indicator function ${\bm 1}(\mc{V}')$ is abbreviated to $1$ here. }
			\label{table:high-order_corrections}
			\centering
			\begin{tabularx}{100mm}{X|c||X|c}
					Diagram					& Equation							& Diagram								& Equation \\ \hline
					$(\circ)$ 					& $\mc{I}_0[0;1]$ 			& $(\bullet)$ 							& $\mc{I}_0[0;e^{is\mc{V}'}-1]$\\ 
					$(\circ \to \circ)$ 		& $\mc{I}_0^2[0;1]$		& $(\circ \to \bullet)$ 			& $\mc{I}_0^2[0;e^{is\mc{V}'}-1]$\\
					$(\circ^n)$		 		& $\mc{I}_0^n[0;1]$		& $(\circ^{n-1} \to \bullet)$	& $\mc{I}_0^n[0;e^{is\mc{V}'}-1]$\\ 
			\end{tabularx}
		\end{table}
		On the basis of the above argument, we generalize the diagrammatic representation toward general multiple-kicks effect. 
		We here introduce the following diagrams: 
		\begin{align}
			(\circ \to \dots \to \circ \to \bullet) &\equiv (\circ^{n-1} \!\to\! \bullet)	\equiv \int_{-\infty}^\infty  \prod_{i=1}^{n} \left[d\Gamma_i {\bm 1}_{[\mc{V}_{i-1},\mc{V}_{i-1}+\mc{Y}_{i}]}^{(i)}\right]\!(e^{is\mc{V}_{n}}\!-\!1) \!=\!\mc{I}_0^{n}[0;e^{is\mc{V}'}\!\!-\!{\bm 1}(\mc{V}')] ,\label{eq:term_n-th_correction1}\\
			(\circ \to \dots \to \circ \to \circ)	 &\equiv (\circ^n)		 	 	 		\equiv \int_{-\infty}^\infty  \prod_{i=1}^{n}   \left[d\Gamma_i {\bm 1}_{[\mc{V}_{i-1},\mc{V}_{i-1}+\mc{Y}_{i}]}^{(i)}\right]=\mc{I}_0^{n}[0;{\bm 1}(\mc{V}')],\label{eq:term_n-th_correction2}
		\end{align}
		where we have introduced $\mc{V}_0\equiv 0$. 
		The diagram $(\circ^n)$ corresponds to the probability that the system is kicked $n$ times during relaxation,
		and the diagram $(\circ^{n-1}\to\bullet)$ corresponds to the effect by the $n$th-kick (the two-dot-dash line in Fig.~{\ref{fig:multiple_kicks_model}(b)})). 
		Using these diagrams, Eq.~{(\ref{eq:relation_exp_multi_kick})} can be rewritten as
		\begin{equation}
			\mu^{n+1}\mc{I}[s;(\circ^{n-1} \to \bullet)] = (\circ^n\to \bullet) - (\circ^n)( \bullet).\label{eq:relation_exp_multi_kick2}
		\end{equation}
		From Eq.~{(\ref{eq:relation_exp_multi_kick2})}, we easily obtain explicit higher-order multiple-kicks processes as 
		\begin{align}
			\mu			\mc{I}^1[s;{\bm 1}(s')] &= (\bullet),\\
			\mu^2	\mc{I}^2[s;{\bm 1}(s')] &= (\circ \to \bullet) - (\circ)(\bullet),\\
			\mu^3	\mc{I}^3[s;{\bm 1}(s')] &= (\circ^2 \to \bullet) - (\circ)(\circ \to \bullet)   - [(\circ^2)- (\circ)^2] (\bullet),\\
			\mu^4	\mc{I}^4[s;{\bm 1}(s')] &= (\circ^3 \to \bullet) - (\circ)(\circ^2 \to \bullet) - [(\circ^2)- (\circ)^2] (\circ \to \bullet) - [(\circ^3) - 2(\circ)(\circ^2) + (\circ)^3](\bullet).
		\end{align}
		We summarize the rules of the diagrams in Table~\ref{table:high-order_corrections}. 
		We note that, when the stopping time diverges as $\tau^*(\mc{Y})\to\infty$ (e.g., the cubic case), 
		the above diagrammatic representation is not valid because some diagrams diverge and termwise integration is not allowed (e.g., the diagram $(\circ)$ diverges for the cubic case). 
		Nevertheless, we note that the formulas~{(\ref{eq:2nd-multi_kick})} and~{(\ref{eq:nth-multi_kick})} are valid even for such cases.

		\subsubsection*{The second-order approximation for the toy model 1 with symmetric Poisson noise}
			\begin{figure}
				\centering
				\includegraphics[width=70mm]{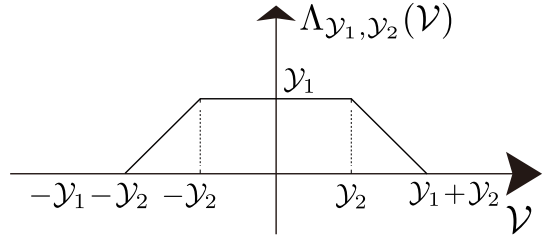}
				\caption{
							Schematic of the trapezoid function defined by Eq.~{(\ref{def:trapezoid})}. 							
						}
				\label{fig:second_Toy1_SPN}
			\end{figure}
			We here explicitly calculate the second-order approximation for the toy model 1 with the symmetric Poisson noise $\mc{W}(\mc{Y})=\lambda[\delta(\mc{Y}-\mc{Y}_0)+\delta(\mc{Y}+\mc{Y}_0)]/2$.
			From Eq.~{(\ref{eq:Toymodel1})}, we obtain the analytic representation of the following diagrams:
			\begin{equation}
				(\circ) 	=  \mu\lambda \mc{Y}_0,\>\>(\bullet) = \mu \lambda \!\left[\frac{\sin{s\mc{Y}_0}}{s}\!-\!\mc{Y}_0\right], \>\>(\circ \to \bullet) = \frac{\mu^2 \lambda^2}{2}\!\left[-\mc{Y}_0^2 + \frac{\cos{s\mc{Y}_0}\!-\!\cos{2s\mc{Y}_0}}{s^2} - \frac{1\!-\!\cos{s\mc{Y}_0}}{s^2}\right].
			\end{equation}
			We then obtain the steady distribution up to the second-order as 
			\begin{equation}
				\mc{P}_{\rm SS}(\mc{V}) = \left[1-\mu\lambda\mc{Y}_0+\frac{\mu^2 \lambda^2\mc{Y}_0^2}{2}\right]\delta(\mc{V}) + 
										\frac{\mu \lambda}{2}\left[1-\mu\lambda\mc{Y}_0\right]{\bm 1}_{[-\mc{Y}_0,\mc{Y}_0]}(\mc{V}) -\frac{\mu^2\lambda^2}{4}\Lambda_{\mc{Y}_0,0}(\mc{V}) + \frac{\mu^2\lambda^2}{4}\Lambda_{\mc{Y}_0,\mc{Y}_0}(\mc{V}),\label{eq:sol_Toymodel1_2nd}
			\end{equation}
			where we have introduced the trapezoid function (see Fig.~{\ref{fig:second_Toy1_SPN}}): 
			\begin{align}
				\Lambda_{\mc{Y}_1,\mc{Y}_2} (\mc{V}) 	
				&\equiv \mc{Y}_1{\bm 1}_{[-\mc{Y}_1-\mc{Y}_2,\mc{Y}_1+\mc{Y}_2]} (\mc{V}) + (\mc{Y}_2-|\mc{V}|)[{\bm 1}_{[-\mc{Y}_1-\mc{Y}_2,-\mc{Y}_2]}(\mc{V})+{\bm 1}_{[\mc{Y}_2,\mc{Y}_1+\mc{Y}_2]}(\mc{V})]\notag\\
				&=\frac{1}{\pi}\int_{-\infty}^\infty ds e^{-is\mc{V}}\frac{\cos{s\mc{Y}_2}-\cos{[s(\mc{Y}_1+\mc{Y}_2)]}}{s^2}. \label{def:trapezoid}
			\end{align}
			We numerically verify the validity of the second-order formula~{(\ref{eq:sol_Toymodel1_2nd})} in Fig.~{\ref{fig:Coulomb_Poisson}(a)}.   
			Note that the convexity of the distribution is violated by the third term on the rhs of Eq.~{(\ref{eq:sol_Toymodel1_2nd})}, which is consistent with the numerical result. 
			We also note that the second-order formula~{(\ref{eq:sol_Toymodel1_2nd})} is valid only for $|\mc{V}|\leq 2\mc{Y}_0$ because the second-order approximation takes in the effect of single and double kicks.

\section{Example: granular motor under dry friction}
	We here apply our formulation to the motion of granular motors, which has been investigated experimentally and theoretically in Refs.~\cite{Eshuis,Gnoli1,Gnoli2,Gnoli3,Talbot,Cleuren,Galvez}. 
	We first explain the setup of the granular motor under dry friction, and introduce the Boltzmann-Lorentz model, which is valid for dilute granular gases~\cite{Brilliantov}. 
	We next show the reduction of the Boltzmann-Lorentz model into the non-Gaussian Langevin equation under dry friction 
	when the mass of the rotor is sufficiently large and the collisions of gases on the rotor are not frequent.
	We then obtain the analytical formula for the steady distribution of the angular velocity of the rotor using the perturbation in terms of the friction coefficient. 
	We numerically verify the validity of our theory from the comparison of the direct Monte Carlo simulation of the Boltzmann-Lorentz model. 
	
	\subsection{Setup}
\begin{figure}
			\centering
			\includegraphics[width=120mm]{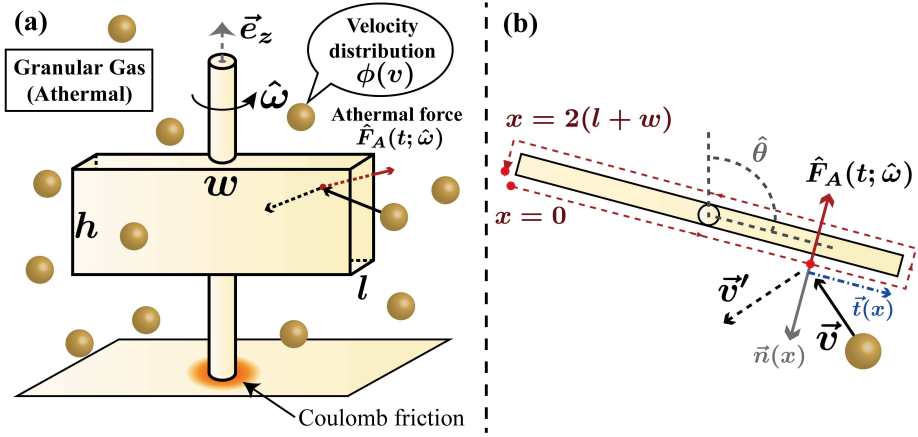}
			\caption{
							(a) Schematic of the granular motor under solid friction. 
							The rotor of cuboid shape ($h\times w\times l$) is located in the granular gas with velocity distribution function (VDF) $\phi(\vec v)$ 
							and is in contact with the solid, where Coulombic friction $\gamma \sgn(\omega)$ is valid. 
							The collisions by granular particles gives the athermal fluctuation $\hat F_A(t;\hat \omega)$. 
							(b) Schematic of the collisional rule for between th rotor and the granular particle. 
							$x$ is the coordinate along the cuboid, where $0\leq x\leq 2(l+w)$
							and $\vec n(x)$ and $\vec t(x)$ are the tangent and horizontal vectors at the point $x$. 
						}
			\label{fig:sc_granular_rotor}
		\end{figure}

		Let us consider a cuboid rotor with mass $M$, inertial moment $I$, height $h$, width $w$, and depth $l$. 
		The rotor is placed in a dilute granular gas and is influenced by dry friction as shown in Fig.~\ref{fig:sc_granular_rotor}(a). 
		The granular gas is externally vibrated and is preserved to be in the non-equilibrium steady state. 
		We assume that the granular gas has an isotropic velocity distribution $\phi(|\vec v|)$ as reported in the experiments\cite{Gnoli1,Gnoli2,Gnoli3}. 
		We also assume that the angular velocity distribution function of the rotor can be described by the Boltzmann-Lorentz equation. 
		We further assume that the dry friction around the $z$-axis can be described by the Amontons-Coulomb law. 
		Then, the dynamics of the rotor are described by the following equation: 
		\begin{equation}
			\frac{\partial P(\om,t)}{\partial t} = \frac{\partial }{\partial \omega}\gamma \sgn (\omega) + \int_{-\infty}^\infty dy[P(\om-y,t)W(\om-y;y)-P(\om,t)W(\om;y)],\label{eq:BL_Master}
		\end{equation}
		where the transition rate is given by
		\begin{equation}
			W(\om;y) = \rho h \int_{0}^{2(l+w)} dx\int_{-\infty}^\infty d\vec v \phi(|\vec v|)\Theta(\Delta \vec V(x)\cdot \vec n(x))|\Delta \vec V(x)\cdot \vec n(x)|\delta(y-\Delta \om(x)).\label{eq:BL_Master_Kernel}
		\end{equation}
		Here, $x$ is the coordinate along the cuboid (see Fig.~\ref{fig:sc_granular_rotor}(b)), $\gamma$ is the friction coefficient, $\vec n(x)$ is the normal unit vector to the surface at the point $x$, 
		$R_I\equiv \sqrt{I/M}$ is the inertia radius, $e$ is the the restitution coefficient between the rotor and the granular particles, and we have introduced the following relations: 
		\begin{equation}
			\vec V(x) \equiv \om \vec e_z\times \vec r(x), \>\>\>g(x)\equiv \frac{\vec r(x)\cdot \vec t(x)}{R_I}, \>\>\>\vec t(x) \equiv \vec e_z \times \vec n(x), \>\>\>\ve \equiv \frac{m}{M},
		\end{equation}
		\begin{equation}
			\Delta \vec V(x) \equiv \vec V(x)-\vec v, \>\>\>\Delta\om(x) \equiv (1+e)\frac{\Delta \vec V(x)\cdot \vec n(x)}{R_I}\frac{\ve g(x)}{1+\ve g^2(x)}.
		\end{equation}
		
	\subsection{Reduction to the non-Gaussian Langevin equation}
		We next take the zero mass-ratio limit $\ve=m/M\to 0$. 
		According to Refs.~\cite{Gnoli1,Gnoli2,Gnoli3}, the characteristics of the dynamics of the rotor depends on whether collisions between the rotor and gases are sufficiently frequent. 
		To characterize the collision frequency, let us introduce the characteristic relaxation time of the rotor caused by Coulombic friction as 
		$\tau_R \equiv mv_0R_I/\gamma I=\ve v_0/\gamma R_I,$
		where $v_0$ is a characteristic granular velocity (e.g., the standard deviation of the velocity for the Maxwellian distribution) and $S\equiv 2(l+w)h$ is the area of the rotor. 
		We also introduce the characteristic collision interval:
		$\tau_C\equiv  (\rho Sv_0)^{-1}.$
		The dynamics of the system are then characterized by the following parameter\footnote{
		We note that the definition of $\beta^{-1}$ is a little different from that in Refs.~\cite{Gnoli1,Gnoli2,Gnoli3}, 
		where $\beta^{-1}$ is defined by $\beta^{-1}\equiv \ve^{1/2}\rho Sv_0^2/\sqrt{2}\pi \gamma R_I$.}:  
		\begin{equation}
			\beta^{-1} \equiv \frac{\tau_R}{\tau_C} = \frac{\ve \rho Sv^2_0}{\gamma R_I}.
		\end{equation}
		For $\beta^{-1} \gg 1$ (the frequent collision limit (FCL)), the collisions are sufficiently frequent and the typical behavior of the angular velocity is well characterized by the Gaussian Langevin equation~\cite{Gnoli3}. 
		For $\beta^{-1} \ll 1$ (the rare collision limit (RCL)), the collisions are so rare that the typical behavior of the system is well described by the independent-kick model~\cite{Gnoli3}. 
		We address the case for $\beta^{-1} \sim 1$ (we call it the occasional collision regime (OCR) in this paper), 
		where the non-Gaussian Langevin equation under Coulombic friction is an appropriate equation in characterizing the dynamical motion of the rotor.
		In the OCR, the friction coefficient $\gamma$ is effectively scaled by $\ve$: 
		\begin{equation}
			\gamma = \ve\tl{\gamma},\label{eq:OCR_scaling}
		\end{equation}
		where we have introduced the scaled frictional coefficient $\tl{\gamma}\equiv \beta \rho S v_0^2/R_I$. 
		The OCR scaling~{(\ref{eq:OCR_scaling})} implies that the dry friction satisfies the conditions (ii) as $\gamma \sgn (\ve \Omega)=\ve \tl{\gamma} \sgn(\Omega)$, where we have introduced the scaled angular velocity $\Omega\equiv \omega/\ve$. 
		Furthermore, the Boltzmann-Lorentz equation satisfies the condition (i) and the dry friction satisfies the condition (iii). 
		In the small $\ve$ limit, then, the master equation is reduced to
		\begin{equation}
			\frac{\partial \mc{P}(\Om,t)}{\partial t} = \frac{\partial }{\partial \Om}\tl{\gamma} \sgn (\Om) + \int_{-\infty}^\infty d\mc{Y}\mc{W}(\mc{Y})[\mc{P}(\Om-\mc{Y},t)-\mc{P}(\Om,t)],\label{eq:BL_SMaster}
		\end{equation}
		where we have introduced the $\Omega$-independent transition rate $\mc{W}(\mc{Y})$ as  
		\begin{equation}
			\mc{W}(\mc{Y}) = \rho h \int_{0}^{2(l+w)} dx\int_{-\infty}^\infty d\vec v \phi(|\vec v|)\Theta(- \vec v\cdot \vec n(x))|\vec v\cdot \vec n(x)|\delta(\mc{Y}-\Delta \Om(x))
		\end{equation}
		with
		\begin{equation}
			\Delta\Om(x) \equiv -(1+e)\frac{\vec v\cdot \vec n(x)}{R_I}g(x).
		\end{equation}
		Equation~{(\ref{eq:BL_SMaster})} is equivalent to the non-Gaussian Langevin equation under Coulombic friction: 
		\begin{equation}
			\frac{d\hat{\Om}}{dt} = -\tl{\gamma}\sgn (\hat \Om) + \hat \xi_{g},\label{eq:NGL_granular_Coulomb}
		\end{equation}
		where the granular noise $\hat \xi_{g}$ is the white non-Gaussian noise whose transition rate is $\mc{W}(\mc{Y})$. 
		We note that the validity of the non-Gaussian Langevin equation~{(\ref{eq:NGL_granular_Coulomb})} has already been experimentally verified in Ref.~\cite{Galvez}.
		We also note that the cumulant generating function of the granular noise is given by
		\begin{equation}
			\Phi(s) = \Phi_l(s) + \Phi_w(s),\label{eq:CGF_granular_noise}
		\end{equation}
		where we have introduced 
		\begin{equation}
			\Phi_p (s) = -\frac{16\pi \rho h R_I^4}{ps^2(1+e)^2}\int_0^\infty \!\!\!\!dv v\phi(v) \left[\cos{\frac{s(1+e)vp}{2R_I^2}} -1 - \frac{s^2(1+e)^2p^2v^2}{8R_I^4}\right]
		\end{equation}
		for an arbitrary real number $p$ (see Appendix.~{\ref{app:CN_cumulant_g}} for the derivation).
		We note the asymptotic tail of the angular velocity under Coulombic friction is given by the exponential form: 
		\begin{equation}
			\mc{P}_{\rm SS}(\Om) \sim e^{-a|\Om|},\label{eq:gmotoro_tail}
		\end{equation}
		where the exponent $a$ is determined by $\tl{\gamma} a=\Phi(-ia)$ as shown in Sec.~\ref{subsec:toy1_Coulomb}. 
		
		We present the explicit forms of the cumulant generating functions for some specific cases.
		Let us first consider the case for Maxwellian velocity distribution function: $\phi(v)=e^{-v^2/2v_0^2}/(2\pi v_0^2)^{3/2}$.
		We note that this condition can be experimentally realized in strong vibration conditions using a specific container in Refs.~\cite{Gnoli3,Olafsen}. 
		In this case, the cumulant generating function is given by 
		\begin{equation}
			\Phi_p (s) = \sqrt{\frac{2}{\pi}}\rho h p v_0 \mc{G}\left(\frac{\Om_p^*s}{\sqrt{2}}\right),
		\end{equation}
		where we have introduced $\Om_p^*\equiv p(1+e)v_0/2R_I^2$ and $\mc{G}(x)\equiv \mc{F}_D(x)/x-1$ with the Dawson function $\mc{F}_D(x)\equiv e^{-x^2}\int_0^x dte^{t^2}$.
		
		We next consider the exponential case: $\phi(v)=e^{-v/v_0}/8\pi v_0^3$. 
		We note that the exponential distribution is also experimentally realized for the weak vibration condition as shown in Ref.~\cite{Olafsen}. 
		We obtain the following form of the cumulant function: 
		\begin{equation}
			\Phi_p (s) = -\frac{\rho hpv_0 \Om_p^{2*}s^2 (5+3\Om_p^{2*}s^2)}{2(1+\Om_p^{2*}s^2)^2}.
		\end{equation}

	\subsection{First-order asymptotic solution}
		Let us analyze the dynamics of the rotor in the RCL condition ($\beta^{-1}\ll 1$). 
		In the RCL, the scaled friction coefficient $\tl{\gamma}$ is  sufficiently large ($\tl{\gamma} \propto 1/\beta^{-1} \gg 1$), 
		and the asymptotic expansion is valid in terms of the inverse of the friction coefficient: $\mu\equiv1/\tl{\gamma}$. 
		We then show the explicit form of the first-order asymptotic formula, which corresponds to the independent-kick model. 
		According to the first-order solution for Coulombic friction~{(\ref{eq:1st_coulomb_toy_dist})}, we obtain the stationary distribution function (see Appendix.~{\ref{app:CN_first_gm}} for the derivation): 
		\begin{equation}
			\mc{P}(\Om) \!=\! \left[1\!-\!\mu c_1\right]\delta(\Om)
			+ \mu\pi \rho hl \int_{|\Om|/F_l}^\infty \!\!\!\!\!\!\!dv v\phi(v)\left[v-\frac{|\Om|}{F_l}\right]^2
			+ \mu\pi \rho hw \int_{|\Om|/F_w}^\infty \!\!\!\!\!\!\!dv v\phi(v)\left[v-\frac{|\Om|}{F_w}\right]^2
			+ O(\mu^2),\label{eq:GM_first}
		\end{equation}
		where we have introduced $F_p\equiv p(1+e)/2R_I^2$ for an arbitrary real number $p$ and 
		\begin{equation}
			c_1 = 2\int_0^\infty d\mc{Y}\mc{W}(\mc{Y})\mc{Y} = \frac{\pi\rho h(1+e)l^2}{3R_I^2}\int_0^\infty dv v^4\phi(v).\label{eq:GM_first_irr}
		\end{equation}
		
		We now study the explicit form of the distribution function~{(\ref{eq:GM_first})} for the following two cases: the Maxwellian and exponential ones. 
		In the Maxwellian case $\phi(v)=e^{-v^2/2v_0^2}/(2\pi v_0^2)^{3/2}$, we obtain 
		\begin{equation}
			\mc{P}(\Om) = \left[1 - \frac{\mu\rho hv_0}{2}(l\Om^*_l+w\Om^*_w)\right]\delta(\Om) + \mc{P}^{(l)}_{\rm smooth}(\Om) + \mc{P}^{(w)}_{\rm smooth}(\Om) +O(\mu^2), \label{eq:BL_dry_1st_1}
		\end{equation}
		where the smooth part $\mc{P}^{(p)}_{\rm smooth}(\Om)$ for an arbitrary real number $p$ is given by
		\begin{equation}
			\mc{P}^{(p)}_{\rm smooth}(\Om) \equiv \frac{\mu\rho hpv_0}{2}\left[\frac{2e^{-|\Om|^2/2\Om_p^{2*}}}{\sqrt{2\pi}} - \frac{|\Om|}{\Om_p^*}{\rm erfc}\left(\frac{1}{\sqrt{2}}\frac{|\Om|}{\Om_p^*}\right)\right]. \label{eq:BL_dry_1st_2}
		\end{equation}
		Here, the complementary error function is defined as ${\rm erfc}(x)\equiv 2\int_x^\infty dte^{-t^2}/\sqrt{\pi}$. 
		In the exponential case $\phi(v)=e^{-v/v_0}/8\pi v_0^3$, we obtain 
		\begin{equation}
			\mc{P}(\Om) = \left[1-\mu \rho hv_0(l\Om^*_l+w\Om^*_w)\right]\delta(\Om) + \mc{P}^{(l)}_{\rm smooth}(\Om) + \mc{P}^{(w)}_{\rm smooth}(\Om) +O(\mu^2), 
		\end{equation}
		where the smooth part $\mc{P}^{(p)}_{\rm smooth}(\Om)$ for an arbitrary real number $p$ is given by
		\begin{equation}
			\mc{P}^{(p)}_{\rm smooth}(\Om) \equiv \frac{\rho h p v_0}{4\tl{\gamma}}\left(3+\frac{|\Om|}{\Om^*_p}\right)e^{-|\Om|/\Om^*_p}.
		\end{equation}

		\subsubsection*{Numerical validation}
			\begin{figure}
				\centering
				\includegraphics[width=150mm]{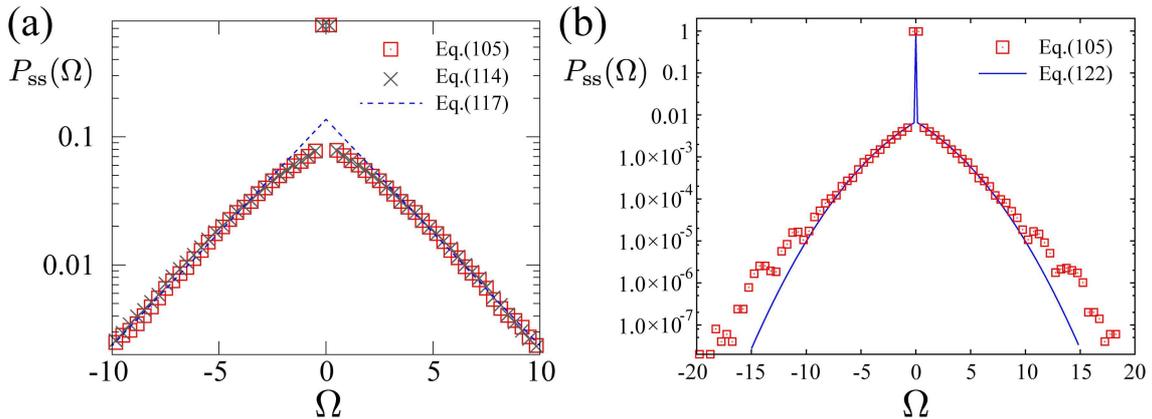}
				\caption{
							(Color online)
							(a) Comparison of the non-Gaussian Langevin equation~{(\ref{eq:NGL_granular_Coulomb})} and the asymptotic tail~{(\ref{eq:gmotoro_tail})} 
								with the direct Monte Carlo simulation~{(\ref{eq:BL_Master})} 
								for $\rho=h=v_0=R_I=I=M=1$, $w=\sqrt{12}$, $m=0.001$, and $\tl{\gamma}=2\sqrt{12}$. 
								The ensemble number and the time discretization for the Monte Carlo simulation are $10^8$ and $3.16\times 10^{-4}$, respectively. 
							(b) Comparison of the independent-kick solution~{(\ref{eq:BL_dry_1st_1})} 
								with the direct Monte Carlo simulation~{(\ref{eq:BL_Master})} 
								for $\rho=h=v_0=R_I=I=M=1$, $w=\sqrt{12}$, $m=0.01$, and $\tl{\gamma}=200$. 
								The ensemble number and the time step for the Monte Carlo simulation are $10^8$ and $10^{-3}$. 
								The first-order asymptotic solution~{(\ref{eq:BL_dry_1st_1})} is only valid in the range $|\Omega| \lesssim 2\Omega_w^*\simeq 7$.
						}
				\label{fig:grotor_C}
			\end{figure}
			We now numerically verify the validity of the non-Gaussian Langevin equation~{(\ref{eq:NGL_granular_Coulomb})} and the asymptotic formula~{(\ref{eq:BL_dry_1st_1})}. 
			We first perform the direct Monte Carlo simulations of the Boltzmann-Lorentz equation~{(\ref{eq:BL_Master})} and the non-Gaussian Langevin equation~{(\ref{eq:NGL_granular_Coulomb})}
			with parameters $\rho=h=v_0=R_I=I=M=1$, $w=\sqrt{12}$, $m=0.001$, and $\tl{\gamma}=2\sqrt{12}$. 
			We note that the above parameters correspond to the OCR; $\beta^{-1}=1$. 
			As shown in Fig.~{\ref{fig:grotor_C}(a)}, the numerical data for the non-Gaussian Langevin equation~{(\ref{eq:NGL_granular_Coulomb})} and the asymptotic tail~{(\ref{eq:gmotoro_tail})} 
			agree with the numerical data for the Boltzmann-Lorentz equation~{(\ref{eq:BL_Master})}. 
			
			We next perform the direct Monte Carlo simulation of the Boltzmann-Lorentz equation~{(\ref{eq:BL_Master})} with parameters $\rho=h=v_0=R_I=I=M=1$, $w=\sqrt{12}$, $m=0.01$, and $\tl{\gamma}=200$, 
			and compare the numerical data with our analytic solution~(\ref{eq:BL_dry_1st_1}). 
			We note that the above parameters correspond to the RCL; $\beta^{-1}\simeq 0.035 \ll 1$. 
			As shown in Fig.~{\ref{fig:grotor_C}(b)}, the first-order asymptotic solution~{(\ref{eq:BL_dry_1st_1})} is valid in the range $|\Omega| \lesssim 2\Omega_w^*\simeq 7$.


\section{Conclusion}
	We derive the non-Gaussian Langevin equation for an arbitrary non-linear friction~{(\ref{eq:NGL_NL_derived})} on the conditions (i) - (iii), 
	and propose an asymptotic connection~{(\ref{eq:NGL_connected})} from the Gaussian to the non-Gaussian Langevin equation. 
	We obtain the full-order asymptotic formula~{(\ref{eq:Solution_Langevin_NLNG})} for the steady distribution function in terms of the inverse of the friction coefficient, 
	and show that the first-order truncation of our formula leads to the independent-kick model.
	Moreover, we show that the higher-order terms directly correspond to the multiple-kicks processes during relaxation by introducing the diagrammatic representations~{(\ref{eq:diagram_ex1})} - {(\ref{eq:term_n-th_correction2})}. 
	We apply our formulation to the granular motor under dry friction, 
	and we systematically show that the dynamics of the rotor can be described by the non-Gaussian Langevin equation~{(\ref{eq:NGL_granular_Coulomb})} and the independent-kick model~{(\ref{eq:GM_first})} 
	under the OCR and RCL conditions, respectively. 
	We numerically verify our formulas for both OCR and RCL conditions. 
	
\begin{acknowledgement}
	We are grateful for the useful discussions between N. Nakagawa and A. Puglisi. 
	A part of the numerical calculations was carried out on SR16000 at YITP in Kyoto University. 
	This work was supported by the JSPS Core-to-Core Program ``Non-equilibrium dynamics of soft matter and information,"
	Grants-in-Aid for the Japan Society for Promotion of Science (JSPS) Fellows (Grant Nos. 24$\cdot$3751 and 26$\cdot$2906), and JSPS KAKENHI Grant Nos. 25287098, and 25800217.
\end{acknowledgement}

\appendix

\section{Relation to the non-equilibrium steady state: Granular rotor under viscous friction}\label{app:gmotor_viscous}
\begin{figure}
		\centering
		\includegraphics[width=120mm]{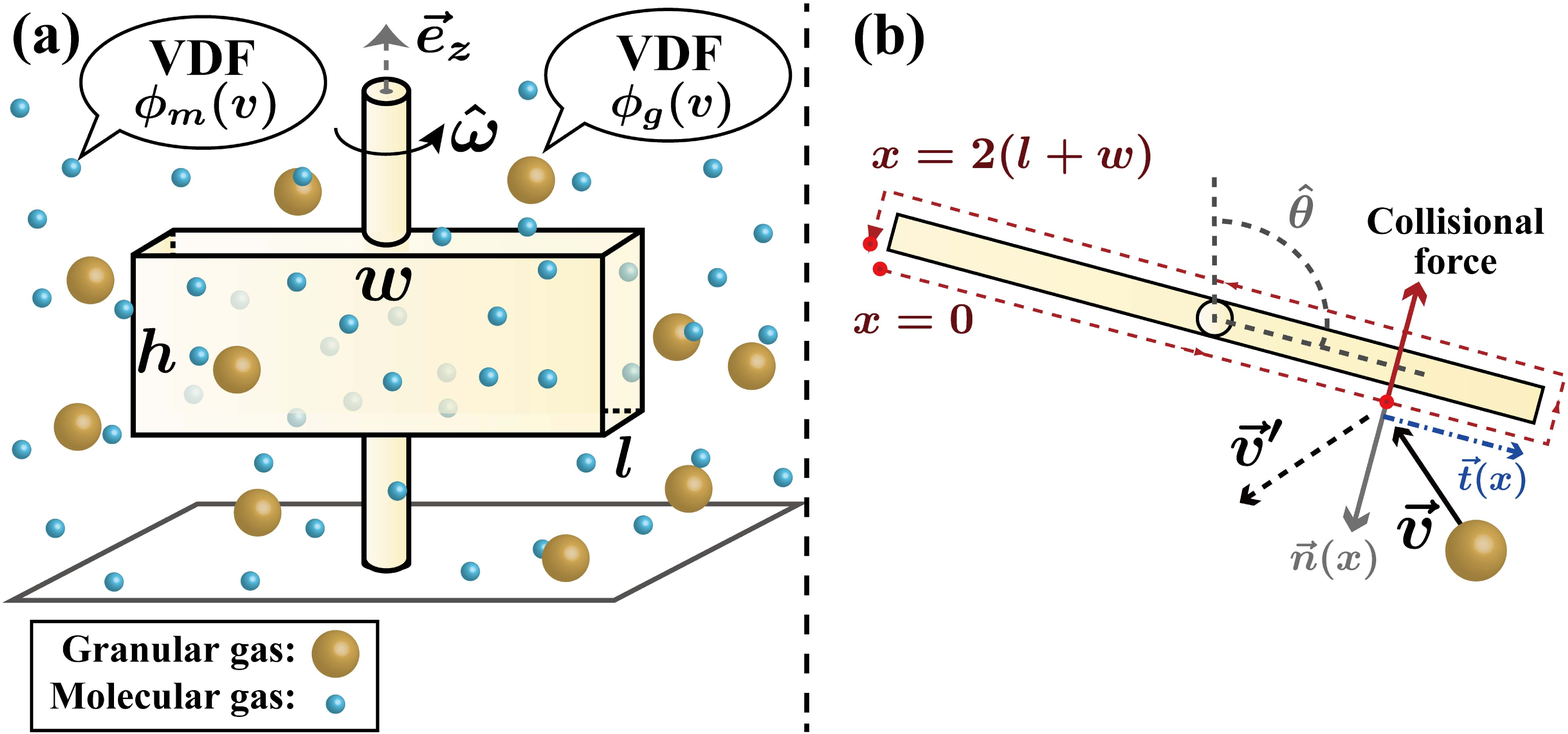}
		\caption{
					(Color online)
					(a) Schematic of rotor associated with the granular and molecular rarefied gases. 
					The rotor is composed of the two cuboids ($h\times w \times l$), and is driven by the collisional impulses by the granular and molecular gas 
					whose velocity distribution functions (VDFs) are $\phi_g(v)$ and $\phi_m(v)$, respectively. 
					The granular and molecular gases are so dilute that their collisional impacts $\hat F_g(t;\omega)$ and $\hat F_m(t;\omega)$ are described by the Boltzmann-Lorentz models with the transition rates $W_g(\omega; y)$ and $W_m(\omega; y)$, respectively. 
					(b) Schematic of the collisional rules between the rotor and a granular (or molecular) particle.  
				}
		\label{fig:gmotor_viscous}
	\end{figure}

	We here study the relation between our formulation and the non-equilibrium steady state through the example of the granular motor under viscous friction. 
	We consider a rotor placed in the granular and molecular rarefied gases characterized by isotropic velocity distributions $\phi_g(|\vec v|)$ and $\phi_m(|\vec v|)$ (see Fig.~{\ref{fig:gmotor_viscous}(a)}). 
	The rotor is cuboid with height $h$, width $w$, and depth $l$. 
	The masses of the rotor, the granular particle, and the molecular particle are $M$, $m_g$, and $m_m$, respectively. 
	For simplicity, we assume that all the restitution coefficients are equal to $1$ and $l=0$. 
	If the granular and molecular gases are sufficiently dilute (i.e., their density $\rho_g$ and $\rho_m$ are sufficiently small), 
	the dynamics of the rotor are governed by the Boltzmann-Lorentz equation as:
	\begin{equation}
		\frac{\partial P(\om,t)}{\partial t} = \sum_{i=g,m} \int_{-\infty}^\infty dy[P(\om-y,t)W_i(\om-y;y)-P(\om,t)W_i(\om;y)],\label{eq:BL_Master2}
	\end{equation}
	\begin{equation}
		W_i(\om;y) = \rho_i h \int_{0}^{2w} dx\int_{-\infty}^\infty d\vec v \phi_i(|\vec v|)\Theta(\Delta \vec V(x)\cdot \vec n(x))|\Delta \vec V(x)\cdot \vec n(x)|\delta(y-\Delta \om_i(x)),\label{eq:BL_Master_Kernel2}
	\end{equation}
	where $x$ is the coordinate along the cuboid (see Fig.~\ref{fig:gmotor_viscous}(b)), $\vec n(x)$ is the normal unit vector to the surface at the point $x$, 
	and we introduce the inertia radius $R_I\equiv \sqrt{I/M}$ and 
	\begin{equation}
		\vec V(x) \equiv \om \vec e_z\times \vec r(x), \>\>\>g(x)\equiv \frac{\vec r(x)\cdot \vec t(x)}{R_I}, \>\>\>\vec t(x) \equiv \vec e_z \times \vec n(x), \>\>\>r_{i} \equiv \frac{m_i}{M}, 
	\end{equation}
	\begin{equation}
		\Delta \vec V(x) \equiv \vec V(x)-\vec v, \>\>\>\Delta\om_i(x) \equiv \frac{\Delta \vec V(x)\cdot \vec n(x)}{R_I}\frac{2r_i g(x)}{1+r_i g^2(x)}.
	\end{equation}
	We here assume that the granular mass ratio $\ve \equiv r_g= m_g/M$ is small.
	Furthermore, we make the following three assumptions: 
	\begin{quote}
		(A1) The masses of the rotor, granular, and molecular particles satisfy the relations $M\gg m_g \gg m_m$. 
				In other words, the mass ratio $r_i$ is scaled as 
				\begin{equation}
					\frac{r_m}{r_g} =O(\ve) \Longleftrightarrow r_m=\ve c_r r_g=\ve^2 c_r,
				\end{equation}
				where $c_r$ is a dimensionless constant independent of $\ve$.
		
		(A2)	The density of the granular gas is much smaller than that of the molecular gas as $\rho_m \gg \rho_g$. 
				In other words, $\rho_g$ is scaled as
				\begin{equation}
					\frac{\rho_g}{\rho_m} = O(\ve^2) \Longleftrightarrow \rho_g = \ve^2 c_\rho \rho_m,
				\end{equation}
				where $c_\rho$ is a dimensionless constant independent of $\ve$. 
				This assumption implies that the collision frequency of the granular particles is much lower than that of the molecular particles. 
 
		(A3)	The velocity distributions $\phi_i(|\vec v|)$ are Maxwellian forms characterized by temperatures $T_i$ for $i=g,m$:
				\begin{equation}
					\phi_i(|\vec v|) = \left(\frac{m_i}{2\pi T_i}\right)^{3/2}\exp{\left[-\frac{m_i|\vec v|^2}{2T_i}\right]}.
				\end{equation}
				Furthermore, the granular temperature $T_g$ is much higher than the molecular temperature $T_m$ as $T_g \gg T_m$. 
				In other words, $T_m$ is scaled with an $\ve$-independent dimensionless constant $c_T$ as
				\begin{equation}
					\frac{T_m}{T_g} = O(\ve) \Longleftrightarrow T_m = \ve c_T T_g.
				\end{equation}
	\end{quote}
	Under the assumptions (A1) - (A3), we use the Kramers-Moyal expansion for the molecular gas:
	\begin{align}
		 \int_{-\infty}^\infty dy[P(\om-y,t)W_m(\om-y;y)-P(\om,t)W_m(\om;y)]
		=&\sum_{n=1}^\infty \frac{(-1)^n\ve^{2n}}{n!}\frac{\partial^n }{\partial \omega^n}\left[K_n(\om)P(\om,t)\right]
	\end{align}
	with the scaled Kramers-Moyal coefficient
	\begin{equation}
		K_n (\om) \equiv \int dy y^nW_m(\om;y)=\rho_m h\int_0^{2w}\!\!dx \int_{-\infty}^\infty d\vec v \phi_m(|\vec v|)\Theta (\Delta \vec V\cdot \vec n)|\Delta \vec V\cdot \vec n|\frac{(2c_r g(x) \Delta \vec V\cdot \vec n)^n}{R_I^n(1+\ve^2 c_rg^2(x))},
	\end{equation}
	where $\phi_m(|\vec v|)$ is independent of $\ve$ as $\phi_m(|\vec v|) = (c_r m_g/2\pi c_TT_g)^{3/2}\exp{[-c_r m_g|\vec v|^2/2c_T T_g]}$. 
	The scaled Kramers-Moyal coefficients are expanded as 
	\begin{equation}
		K_n (\om) = \sum_{k=0}^\infty \frac{K_{n;(k)}^*}{k!}\om^{k},
	\end{equation}
	where $K_{1;(0)}^*=0$ and $K_{1;(1)}^*\neq 0$. 
	Introducing the scaled variables 
	\begin{equation}
		\tau \equiv \ve^2t,\>\>\>\> \Om \equiv \frac{\om}{\ve},
	\end{equation}
	we obtain the scaled master equation~{(\ref{eq:BL_Master2})} as 
	\begin{align}
		\frac{\partial \mc{P}(\Om,\tau)}{\partial \tau} &= \left[\sum_{k=0}^\infty\frac{\ve^k}{k!}\left\{\frac{K_{1;(k+1)}^*}{k+1}\frac{\partial }{\partial \Om}\Om^{k+1} 
																								+\frac{K_{2;(k)}^*}{2}\frac{\partial }{\partial \Om}\Om^{k} \right\}
																								+ \sum_{n=3}^\infty \sum_{k=0}^\infty \frac{(-1)^n\ve^{n+k-2}K_{n;(k)}^*}{n!k!}\frac{\partial^n }{\partial \Om^n}\Om^k\right]\mc{P}(\Om,\tau)\notag\\
		&+\int_{-\infty}^\infty d\mc{Y}[\mc{P}(\Om-\mc{Y},\tau)\tl{W}_g(\Om-\mc{Y};\mc{Y})-\mc{P}(\Om,\tau)\tl{W}_g(\Om;\mc{Y})],\label{eq:gm_rotor_trans}
	\end{align}
	\begin{equation}
		\tl{W}_g(\Om;\mc{Y}) = c_\rho \rho_m h \int_{0}^{2w} dx\int_{-\infty}^\infty d\vec v \phi_g(|\vec v|)\Theta(\Delta \vec {\tl{V}}(x)\cdot \vec n(x))|\Delta \vec{\tl{V}}(x)\cdot \vec n(x)|\delta(\mc{Y}-\Delta \Om_g(x)),
	\end{equation}
	where $\phi_g(|\vec v|)$ is independent of $\ve$ as $\phi_g(|\vec v|) = (m_g/2\pi T_g)^{3/2}\exp{[-m_g|\vec v|^2/2 T_g]}$ and we introduce 
	\begin{equation}
		\vec{\tl{V}}(x) \equiv \Om \vec e_z\times \vec r(x), \>\>\>\Delta \vec{\tl{V}}(x) \equiv \ve \vec{\tl{V}}(x)-\vec v, \>\>\> \Delta\Om_g(x) \equiv \frac{\Delta \vec{\tl{V}}(x)\cdot \vec n(x)}{R_I}\frac{2g(x)}{1+\ve g^2(x)}.
	\end{equation}
	In the limit $\ve \to 0$, Eq.~{(\ref{eq:gm_rotor_trans})} is reduced to
	\begin{equation}
		\frac{\partial \mc{P}(\Om, \tau)}{\partial \tau} 	= \gamma \left[\frac{\partial }{\partial \Om}\Om + \frac{\mc{T}_m}{\mc{I}}\frac{\partial^2 }{\partial \Om^2}\right]\mc{P}(\Om, \tau)
															+ \int_{-\infty}^\infty \mc{W}(\mc{Y})[\mc{P}(\Om-\mc{Y}, \tau)-\mc{P}(\Om, \tau)],\label{app:eq:smaster}
	\end{equation}
	\begin{equation}
		\mc{W}(\mc{Y}) = 2c_\rho \rho_m h \int_{0}^{w} dx\int_{-\infty}^\infty d\vec v \phi_g(|\vec v|)\Theta(- \vec v\cdot \vec n(x))|\vec v\cdot \vec n(x)|\delta(\mc{Y}-\Delta \Om(x)),
	\end{equation}
	where we have used 
	\begin{equation}
		 K^*_{1;(1)} = \frac{hw^3\rho_m}{3R_I^2}\sqrt{\frac{2c_r c_T T_g}{\pi m_g}} \equiv \gamma,\>\>\> K_{2;(0)}^* = \frac{2hw^3 \rho_m c_T T_g}{3R_I^4m_g}\sqrt{\frac{2c_r c_T T_g}{\pi m_g}} =\frac{2\gamma \mc{T}_m}{\mc{I}} 
	\end{equation}
	with $\mc{T}_m\equiv c_T T_g$ and $\mc{I}\equiv m_gR_I^2$. 
	Equation~{(\ref{app:eq:smaster})} is equivalent to the non-Gaussian Langevin equation
	\begin{equation}
		\frac{d\hat{\Om}}{d\tau} = -\gamma \hat \Om + \sqrt{2\gamma \mc{T}_m/\mc{I}}\hat \xi_{\rm G} + \hat \xi_{\rm NG},\label{app:eq:NGL_gmotor_v}
	\end{equation}
	where $\hat \xi_{\rm NG}$ is characterized by the transition rate $\mc{W}(\mc{Y})$. 
	As can be seen in the assumption (A3), the non-Gaussian Langevin equation~{(\ref{app:eq:NGL_gmotor_v})} is only valid 
	for system connected with two reservoirs characterized by extremely different temperatures. 
	If there is no temperature difference (i.e., the system is in equilibrium condition as $T_g=T_m$), the non-Gaussian Langevin equation~{(\ref{app:eq:NGL_gmotor_v})} does not appear. 
	We also note that there exists a energy current from the granular to the molecular gas. 	

\section{Check of the solution~{(\ref{eq:iterative_IE})} of the integral equation~{(\ref{eq:iterative_IE_trans})}}\label{app:CN:IntegralEq}
	In this appendix, we check that the solution~{(\ref{eq:iterative_IE})} satisfies the integral equation~{(\ref{eq:iterative_IE_trans})} on the condition that $\tl{a}_n(0)=0$. 
	We note that Eq.~{(\ref{eq:iterative_IE_trans})} belongs to the class of the first-kind Fredholm integral equations with convolution kernels~\cite{Arfken}. 
	Substituting the solution~{(\ref{eq:iterative_IE})} into the left hand-side of Eq.~{(\ref{eq:iterative_IE_trans})}, we obtain 
	\begin{align}
		\frac{1}{2\pi}\int_{-\infty}^\infty du \tl{f}(s-u)\tl{a}_{n+1}(u) 	
			&= \frac{1}{2\pi}\int_{-\infty}^\infty du \tl{f}(s-u)\mc{I}[u;\tl{a}_n(s')]\notag\\
			&= \frac{1}{(2\pi)^2}\int_{-\infty}^\infty du \tl{f}(s-u)\int_{-\infty}^\infty \frac{d\mc{V}(e^{iu\mc{V}}-1)}{f(\mc{V})}\int_{-\infty}^\infty ds'e^{-is'\mc{V}}\frac{\Phi(s')}{is'}\tl{a}_n(s')\notag\\
			&= \frac{1}{(2\pi)^2}\int_{-\infty}^\infty \frac{d\mc{V}}{f(\mc{V})}\int_{-\infty}^\infty du \tl{f}(s-u)(e^{iu\mc{V}}-1)\int_{-\infty}^\infty ds'e^{-is'\mc{V}}\frac{\Phi(s')}{is'}\tl{a}_n(s')\notag\\
			&= \frac{1}{2\pi}\int_{-\infty}^\infty \frac{d\mc{V}}{f(\mc{V})}f(\mc{V})e^{is\mc{V}}\int_{-\infty}^\infty ds'e^{-is'\mc{V}}\frac{\Phi(s')}{is'}\tl{a}_n(s')\notag\\
			&= \int_{-\infty}^\infty ds'\delta(s-s')\frac{\Phi(s')}{is'}\tl{a}_n(s') = \frac{\Phi(s)}{is}\tl{a}_n(s),
	\end{align}
	where we have used the relation $\int_{-\infty}^\infty du\tl{f}(s-u)=f(0)=0$ in the third line. 
	We note that the solution~{(\ref{eq:iterative_IE})} satisfies the condition for the conservation of the probability $\tl{a}_n(0)=0$. 
	Equation~{(\ref{eq:iterative_IE})} is then the solution of the integral equation~{(\ref{eq:iterative_IE_trans})}. 
	
\section{Derivation of the asymptotic tail~{(\ref{eq:asymptotic_tail_cubic})} for the cubic friction}\label{app:CN:tail_cubic}
	We here check that the explicit form of the asymptotic tail~{(\ref{eq:asymptotic_tail_cubic})} for the cubic friction. 
	We first assume that $|d\tl{P}/ds| \gg |d^3\tl{P}/ds^3|$ for $s\to \infty$. 
	Using the method of dominant balance, we obtain 
	\begin{equation}
		\frac{d\tl{P}(s)}{ds} \simeq \frac{\mu\Phi(s)}{s}\tl{P}(s) \Longrightarrow \tl{P}(s) \simeq \exp\left[\mu\int_0^s \frac{\Phi(s')}{s'}\right] = \exp\left[2\mu \int_0^\infty d\mc{Y}\mc{W}(\mc{Y})\int_0^\mc{Y} d\mc{V}\frac{\cos{s\mc{V}}-1}{\mc{V}}\right], \label{app:asymp_cubic_sol}
	\end{equation}
	where we have used the relation $\int_0^s ds'(\cos{s'\mc{Y}}-1)/s' = \int_0^{\mc{Y}} d\mc{V}(\cos{s\mc{V}}-1)/\mc{V}$. 
	Note that $d\tl{P}(s)/ds$ and $d^3\tl{P}(s)/ds^3$ decay for $s\to \infty$ as 
	\begin{equation}
		\frac{1}{\tl{P}(s)}\frac{d\tl{P}(s)}{ds} \sim -\frac{\mu \lambda^*}{s} + o(s^{-1}),\>\>\> \frac{1}{\tl{P}}\frac{d^3\tl{P}(s)}{ds^3} \sim -\frac{\mu^3\lambda^{*3}}{s} + o(s^{-1}),\label{app:tail_cubic_transient1}
	\end{equation}
	where we have introduced $\lambda^* \equiv 2\int_{0}^\infty d\mc{Y}\mc{W}(\mc{Y})$ and used the relation $\lim_{s\to \infty}\Phi(s) = -\lambda^*$.
	Equation~{(\ref{app:tail_cubic_transient1})} ensures the consistency of the assumption $|d\tl{P}/ds| \gg |d^3\tl{P}/ds^3|$ for $s\to \infty$. 
	We note that the solution~{(\ref{app:asymp_cubic_sol})} asymptotically behaves as
	\begin{equation}
		\tl{P}(s) \simeq \exp\left[-2\mu\int_0^\infty d\mc{Y}\mc{W}(\mc{Y}){\rm Cin}(s\mc{Y})\right] \sim \exp\left[-2\mu \int_0^\infty d\mc{Y}\mc{W}(\mc{Y})\log{s}\right]=|s|^{-\mu\lambda^*},
	\end{equation}
	where we have used the asymptotic form of the cosine integral ${\rm Cin}(x)\equiv \int_0^x dt(1-\cos{t})/t \sim \log{x}$ for $x\to \infty$.
	
\section{Check of the renormalized solution~{(\ref{eq:asymp_1st_cubic_toy_dist})}}\label{app:CN:renormalized}
	In this appendix, we check whether the postulated expression~{(\ref{eq:asymp_1st_cubic_toy_dist})} satisfies the ordinary differential equation~{(\ref{eq:ODE_cubic_toy})} by the direct substitution. 
	For simplicity, we assume that $\mc{W}(\mc{Y})$ is an $L^2$-function, where $\Phi(s)$ is an bounded function as $|\Phi(s)| \leq \lambda^*$ with $\lambda^* \equiv 2\int_{0}^\infty d\mc{Y}\mc{W}(\mc{Y})$. 
	We note that $\Phi(s)/s$ is also a bounded function because $\Phi(s)/s$ is regular at $s=0$ as $\lim_{s\to 0}\Phi(s)/s = 0$ due to the symmetry $\Phi(s)=\Phi(-s)$. 
	We also note that $\Phi(s)/s$ behaves as $\Phi(s)/s = O(s^{-1})$ in the limit $s\to \infty$. 
	The differential equation~{(\ref{eq:ODE_cubic_toy})} then has only two singular points at $s=\pm \infty$.
	We here introduce the following quantities: 
	\begin{equation}
		Q_1(s) \equiv 2\int_0^\infty d\mc{Y}\mc{W}(\mc{Y})\int_0^\mc{Y} d\mc{V}\frac{\cos{s\mc{V}}-1}{\mc{V}(1+\mc{V}^2)}, \>\>\>Q_2(s) \equiv \frac{dQ_1}{ds} = -2\int_0^\infty d\mc{Y}\mc{W}(\mc{Y})\int_0^\mc{Y} d\mc{V}\frac{\sin{s\mc{V}}}{1+\mc{V}^2},
	\end{equation}
	\begin{equation}
		Q_3(s) \equiv \frac{dQ_2}{ds} = -2\int_0^\infty d\mc{Y}\mc{W}(\mc{Y})\int_0^\mc{Y} d\mc{V}\frac{\mc{V}\cos{s\mc{V}}}{1+\mc{V}^2}, \>\>\>Q_4(s) \equiv \frac{dQ_3}{ds} = 2\int_0^\infty d\mc{Y}\mc{W}(\mc{Y})\int_0^\mc{Y} d\mc{V}\frac{\mc{V}^2\sin{s\mc{V}}}{1+\mc{V}^2}.
	\end{equation}
	For these quantities, the following relations hold:
	\begin{equation}
		\frac{d}{ds}e^{\mu Q_1(s)} 		= \mu Q_2 e^{\mu Q_1(s)},\>\>\>
		\frac{d^3}{ds^3}e^{\mu Q_1(s)}	= \left[\mu Q_4 + 3\mu^2 Q_2Q_3+\mu^3 Q_2^3\right]e^{\mu Q_1(s)}.\label{app:transient_1}
	\end{equation}
	Then, we obtain 
	\begin{align}
		\left(\frac{d}{ds}-\frac{d^3}{ds^3}\right)e^{\mu Q_1(s)}	&= -\left(2\mu \int_0^\infty d\mc{Y}\mc{W}(\mc{Y})\int_0^{\mc{Y}}\sin{s\mc{V}}\right)e^{\mu Q_1(s)} + \left[3\mu^2 Q_2Q_3+\mu^3 Q_2^3\right]e^{\mu Q_1(s)}\notag\\
															&= \left(2\mu \int_0^\infty d\mc{Y}\mc{W}(\mc{Y})(\cos{s\mc{Y}}-1)\right)e^{\mu Q_1(s)} + \left[3\mu^2 Q_2Q_3+\mu^3 Q_2^3\right]e^{\mu Q_1(s)}\notag\\
															&= \frac{\mu \Phi(s)}{s}e^{\mu Q_1(s)} + \left[3\mu^2 Q_2Q_3+\mu^3 Q_2^3\right]e^{\mu Q_1(s)}.\label{app:transient_2}
	\end{align}
	Note that $Q_2$ and $Q_3$ are bounded as
	\begin{equation}
		|Q_2(s)| \leq 2\int_0^\infty d\mc{Y}\mc{W}(\mc{Y})\int_0^\mc{Y} d\mc{V}\left|\frac{\sin{s\mc{V}}}{1+\mc{V}^2}\right| \leq 2E,\>\>\> |Q_3| \leq 2\int_0^\infty d\mc{Y}\mc{W}(\mc{Y})\int_0^\mc{Y} d\mc{V}\left|\frac{\mc{V}^2\sin{s\mc{V}}}{1+\mc{V}^2}\right| \leq E,
	\end{equation}
	where $E\equiv \int_0^\infty d\mc{Y}\mc{W}(\mc{Y})\mc{Y}>0$. 
	Furthermore, $\Phi(s)/s$, $Q_2$, and $Q_3$ decay for $s\to \infty$ as
	\begin{align}
		\frac{\Phi(s)}{s} 	&=   \int_0^\infty d\mc{Y}\mc{W}(\mc{Y})\frac{\cos{s\mc{Y}}-1}{s} \simeq - \frac{\lambda^*}{2s} +O(s^{-2}), \\
		Q_2(s) 				&= -2\int_0^\infty d\mc{Y}\mc{W}(\mc{Y})\int_0^\mc{Y} d\mc{V}\frac{\sin{s\mc{V}}}{1+\mc{V}^2} \simeq -2\int_0^\infty d\mc{Y}\mc{W}(\mc{Y})\int_0^\infty d\mc{V}\frac{\sin{s\mc{V}}}{1+\mc{V}^2} = -\frac{\lambda^*}{s} + O(s^{-3}),\\
		Q_3(s)    			&= -2\int_0^\infty d\mc{Y}\mc{W}(\mc{Y})\int_0^\mc{Y} d\mc{V}\frac{\mc{V}\cos{s\mc{V}}}{1+\mc{V}^2} \simeq -2\int_0^\infty d\mc{Y}\mc{W}(\mc{Y})\int_0^\infty d\mc{V}\frac{\mc{V}\cos{s\mc{V}}}{1+\mc{V}^2} = \frac{\lambda^*}{s^2} + O(s^{-4}), 
	\end{align}
	where we have used the Riemann-Lebesgue lemma as $\lim_{s\to \infty}\int_0^\infty \mc{W}(\mc{Y}) \cos{s\mc{V}}=0$. 
	This implies that the second term on the rhs of Eq.~{(\ref{app:transient_2})} is negligible compared with the first term for $s\to \infty$ and 
	that $\tl{P}(s)=e^{Q_1(s)/\gamma}$ uniformly satisfies Eq.~{(\ref{eq:ODE_cubic_toy})} up to the first-order. 
	We also note asymptotic relations as
	\begin{align}
		\left|e^{-\mu Q_1(s)}\left[\left(\frac{d}{ds}-\frac{d^3}{ds^3}\right)e^{\mu Q_1(s)} - \frac{\mu \Phi(s)}{s}e^{\mu Q_1(s)}\right] \right| &\leq 6\mu^2E^2 + 8\mu^3 E^3,\\
		\left|e^{-\mu Q_1(s)}\left[\left(\frac{d}{ds}-\frac{d^3}{ds^3}\right)e^{\mu Q_1(s)} - \frac{\mu \Phi(s)}{s}e^{\mu Q_1(s)}\right] \right| &= \frac{-3\mu^2\lambda^{2*}}{ s^3} + \frac{\mu^3\lambda^{2*}}{ s^2} + o(s^{-4}).\>\>\> (s\to \infty)
	\end{align}
	
\section{Cumulant generating function of the granular noise~{(\ref{eq:CGF_granular_noise})}}\label{app:CN_cumulant_g}
	In this appendix, we derive the explicit form of the cumulant generating function of the granular noise~{(\ref{eq:CGF_granular_noise})}. 
	The cumulant generating function $\Phi(s)$ can be transformed as 
	\begin{align}
		\Phi(s) &= \int_{-\infty}^\infty d\mc{Y}\mc{W}(\mc{Y})(e^{is\mc{Y}}-1)\notag\\
				&= \rho h\int_{-\infty}^\infty d\mc{Y} \int_{0}^{2(l+w)}dx \int_{-\infty}^\infty d\vec v (e^{is\mc{Y}}-1)\phi(|\vec v|)\Theta(-\vec v\cdot \vec n(x))|\vec v\cdot \vec n(x)|\delta\left[\mc{Y}+(1+e)\frac{\vec v\cdot \vec n(x)}{R_I}g(x)\right]\notag\\
				&= \rho h\int_{0}^{2(l+w)}dx \int_{-\infty}^\infty d\vec v \phi(|\vec v|)\Theta(-\vec v\cdot \vec n(x))|\vec v\cdot \vec n(x)|(e^{-is(1+e)(\vec v \cdot \vec n(x))g(x)/R_I}-1).
	\end{align}
	Introducing the representation of the polar coordinate system $(v,\theta',\psi)$, 
	we obtain
	\begin{align}
		\Phi(s) &= \rho h\int_{0}^{2(l+w)}\!\!\!\! dx \int_0^\infty \!\!\!\!dv \int_0^{2\pi}\!\!\!\!d\theta' \int_0^\pi \!\!\!d\psi v^2\sin{\psi}  \phi(v)\Theta(-v\cos{\psi})|v\cos{\psi}|(e^{-is(1+e)v\cos{\psi}g(x)/R_I}-1)\notag\\
				&= -\rho h\int_{0}^{2(l+w)}\!\!\!\! dx \int_0^\infty \!\!\!\!dv \int_0^{2\pi}\!\!\!\!d\theta' \int_{\pi/2}^\pi \!\!\!d\psi v^3  \phi(v)\sin{\psi}\cos{\psi}(e^{-i[s(1+e)vg(x)/R_I]\cos{\psi}}-1)\notag\\
				&= 4\pi \rho h\int_{0}^{(l+w)}\!\!\!\! dx \int_0^\infty \!\!\!\!dv \int_0^1 d\chi v^3\phi(v)\chi(e^{is(1+e)vg(x)\chi/R_I}-1)-\notag\\
				&= \Phi_l(s) + \Phi_w(s),
	\end{align}
	where we have introduced $\chi = -\cos\psi$ and 
	\begin{align}
		\Phi_l (s) &= 4\pi \rho h\int_{0}^{l} \>\>dx \int_0^\infty \!\!\!\!dv \int_0^1 d\chi v^3\phi(v)\chi(e^{is(1+e)vg(x)\chi/R_I}-1), \label{app:cumulant_l}\\
		\Phi_w (s) &= 4\pi \rho h\int_{l}^{(l+w)}\!\!\!\!\!\!\!\! dx \int_0^\infty \!\!\!\!dv \int_0^1 d\chi v^3\phi(v)\chi(e^{is(1+e)vg(x)\chi/R_I}-1).
	\end{align}
	Substituting $g(x)=(x-l/2)/R_I$ for $0\leq x\leq l$ into Eq.~{(\ref{app:cumulant_l})}, we obtain
	\begin{align}
		\Phi_l (s) 	&= 4\pi \rho h\int_{0}^{l}dx \int_0^\infty \!\!\!\!dv \int_0^1 d\chi v^3\phi(v)\chi(e^{is(1+e)v(x-l/2)\chi/R^2_I}-1) \notag\\
					&= 4\pi \rho h\int_0^\infty \!\!\!\!dv v^3\phi(v) \int_0^1 d\chi \chi \left[ \frac{2\sin{\frac{s(1+e)v\chi l}{2R_I^2}}}{s(1+e)v\chi /R_I^2}- l\right] \notag\\
					&= 4\pi \rho h\int_0^\infty \!\!\!\!dv v^3\phi(v) \left[ \frac{1-\cos{\frac{s(1+e)v l}{2R_I^2}}}{[s(1+e)v /2R_I^2]^2l}- \frac{l}{2}\right]\notag\\
					&= -\frac{16\pi \rho h R_I^4}{ls^2(1+e)^2}\int_0^\infty \!\!\!\!dv v\phi(v) \left[\cos{\frac{s(1+e)v l}{2R_I^2}} -1 + \frac{s^2(1+e)^2l^2v^2}{8R_I^4}\right].
	\end{align}
	We similarly obtain
	\begin{equation}
		\Phi_w (s) = -\frac{16\pi \rho h R_I^4}{ws^2(1+e)^2}\int_0^\infty \!\!\!\!dv v\phi(v) \left[\cos{\frac{s(1+e)v w}{2R_I^2}} -1 + \frac{s^2(1+e)^2w^2v^2}{8R_I^4}\right].
	\end{equation}

\section{First order solution of the angular velocity's distribution for the granular motor}\label{app:CN_first_gm}
	We here write the explicit derivation of Eq.~{(\ref{eq:GM_first})}. 
	From Eqs.~{(\ref{eq:1st_coulomb_toy_dist})} and~{(\ref{eq:BL_Master_Kernel})}, we obtain 
	\begin{align}
		\mc{P}_{\rm SS} (\Om) 	&= \left[1-\frac{c_1}{\tl{\gamma}}\right]\delta(\Om) + \frac{ \rho h}{\tl{\gamma}}\int_0^\infty d\mc{Y}{\bm 1}_{[-\mc{Y},\mc{Y}]}(\Om)\int_0^{2(l+w)}dx \int d\vec v\phi(|\vec v|)\Theta (-\vec v\cdot \vec n(x))|\vec v\cdot \vec n(x)|\delta(\mc{Y}-\Delta\Om (x))\notag\\
								&= \left[1-\frac{c_1}{\tl{\gamma}}\right]\delta(\Om) + \frac{2\rho h}{\tl{\gamma}}\int_0^\infty d\mc{Y}{\bm 1}_{[-\mc{Y},\mc{Y}]}(\Om)\int_0^{ (l+w)}dx \int d\vec v\phi(|\vec v|)\Theta (-\vec v\cdot \vec n(x))|\vec v\cdot \vec n(x)|\delta(\mc{Y}-\Delta\Om (x)),\label{app:dist_first_transient1}
	\end{align}
	where we have introduced $c_1 = \int_0^\infty d\mc{Y}\mc{W}(\mc{Y})\mc{Y}$.
	Here we calculate the following integral: 
	\begin{align}
		 &\int_0^\infty d\mc{Y}{\bm 1}_{[-\mc{Y},\mc{Y}]}(\Om)\int_0^{l}dx \int d\vec v\phi(|\vec v|)\Theta (-\vec v\cdot \vec n(x))|\vec v\cdot \vec n(x)|\delta(\mc{Y}-\Delta\Om (x))\notag\\
		=&\int_0^\infty d\mc{Y}{\bm 1}_{[-\mc{Y},\mc{Y}]}(\Om)\int_0^{l}dx \int_0^\infty \!\!\!\!dv\int_0^{2\pi}\!\!\!\!d\theta'\int_0^\pi \!\!\!\!d\psi v^3 \sin\psi(-\cos\psi) \phi(v)\Theta (-v\cos{\psi})\delta\left[\mc{Y}+\frac{(1+e)(x-l/2)v\cos\psi}{R_I^2}\right]\notag\\
		=&2\pi \int_{l/2}^{l}dx \int_0^\infty dv\int_{0}^1 d\chi \chi v^3 \phi(v){\bm 1}_{[-\mc{Y}^{\dagger}(x,v,\chi),\mc{Y}^{\dagger}(x,v,\chi)]}(\Om), \label{app:transient_first_order}
	\end{align}
	where we have introduced the polar coordinate $(v,\theta',\psi)$, the flight distance $\mc{Y}^\dagger(x,v,\psi) \equiv (1+e)(x-l/2)\chi v/R_I^2$, and the variable transformation $\chi = -\cos\psi$, and have used $g(x)=(x-l/2)/R_I$ for $0\leq x \leq l$. 
	We remark that
	\begin{equation}
		|\Om| \leq \mc{Y}^{\dagger}(x,v,\chi) \Longleftrightarrow \frac{R_I^2 |\Om|}{(1+e)(x-l/2)\chi }\leq v.
	\end{equation}
	We then rewrite the rhs of Eq.~{(\ref{app:transient_first_order})} into the following form:
	\begin{align}
		 2\pi \int_{l/2}^{l}dx \int_0^\infty dv\int_{0}^1 d\chi \chi v^3 \phi(v){\bm 1}_{[-\mc{Y}^{\dagger}(x,v,\chi),\mc{Y}^{\dagger}(x,v,\chi)]}(\Om)
		&=2\pi \int_0^{l/2} dx' \int_0^1 d\chi \chi\int_{R_I^2|\Om|/(1+e)x'\chi}^\infty \!\!dv v^3 \phi(v)\notag\\
		&=\frac{\pi l}{2}\int_{2R_I^2|\Om|/(1+e)l}^\infty dv v\phi(v)\left[v-\frac{2R_I^2|\Om|}{(1+e)l}\right]^2,
	\end{align}
	where we have introduced the variable transformation $x'=x-l/2$ and used the identity for an arbitrary positive number $c$ as 
	\begin{equation}
		\int_0^{l/2}dx' \int_0^1 d\chi \chi \int_{c/\chi x'}^{\infty} dv v^3\phi(v) = \frac{l}{4}\int_{2c/l}^\infty dv v\phi(v)\left[v-\frac{2c}{l}\right]^2.
	\end{equation}
	Similarly, we obtain 
	\begin{align}
		 &\int_0^\infty d\mc{Y}{\bm 1}_{[-\mc{Y},\mc{Y}]}(\mc{V})\int_{l}^{(l+w)}dx \int d\vec v\phi(|\vec v|)\Theta (-\vec v\cdot \vec n(x))|\vec v\cdot \vec n(x)|\delta(\mc{Y}-\Delta\Om (x))\notag\\
		=&\frac{\pi w}{2}\int_{2R_I^2|\Om|/(1+e)w}^\infty dv v\phi(v)\left[v-\frac{2R_I^2|\Om|}{(1+e)w}\right]^2.
	\end{align}
	We then obtain 
	\begin{align}
		 &\frac{\rho h}{\tl{\gamma}}\int_0^\infty d\mc{Y}{\bm 1}_{[-\mc{Y},\mc{Y}]}(\Om)\int_0^{2(l+w)}dx \int d\vec v\phi(|\vec v|)\Theta (-\vec v\cdot \vec n(x))\delta(\mc{Y}-\Delta\Om (x))\notag\\
		=&\frac{\pi \rho hl}{\tl{\gamma}}\int_{|\Om|/F_l}^\infty \!\!\!\!\!\!\!dv v\phi(v)\left[v-\frac{|\Om|}{F_l}\right]^2
			+ \frac{\pi \rho hw}{\tl{\gamma}}\int_{|\Om|/F_w}^\infty \!\!\!\!\!\!\!dv v\phi(v)\left[v-\frac{|\Om|}{F_w}\right]^2,\label{app:dist_first_transient2}
	\end{align}
	where we have used $F_p\equiv p(1+e)/2R_I^2$ for an arbitrary real number $p$.
	From Eqs.~{(\ref{app:dist_first_transient1})} and~{(\ref{app:dist_first_transient2})}, we obtain Eq.~{(\ref{eq:GM_first})}.
	We also obtain the explicit form of $c_1$ as follows:
	\begin{equation}
		c_1 = 2\int_0^\infty d\mc{Y}\mc{W}(\mc{Y})\mc{Y}
				= 4\rho h\int_0^\infty d\mc{Y}\mc{Y}\int_0^{(l+w)} dx \int_{-\infty}^{\infty} d\vec v \phi(|\vec v|)\Theta(-\vec v\cdot \vec n(x))\delta(\mc{Y}-\Delta \Om(x)).\label{app:w_tot_transient1}
	\end{equation}
	We here calculate the following integral as 
	\begin{align}
		 &\int_0^\infty d\mc{Y}\mc{Y}\int_0^{l} dx \int_{-\infty}^{\infty} d\vec v \phi(|\vec v|)\Theta(-\vec v\cdot \vec n(x))|\vec v\cdot \vec n(x)|\delta(\mc{Y}-\Delta \Om(x))\notag\\
		=&2\pi \int_0^\infty d\mc{Y}\mc{Y}\int_0^{l} dx \int_0^{\infty} dvv^2 \int_{0}^{\pi}d\psi \sin\psi \phi(v)\Theta(-v\cos{\psi})|v\cos\psi|\delta\left(\mc{Y}-\frac{(1+e)(x-l/2)v\cos\psi}{R_I^2}\right)\notag\\
		=&\frac{2\pi (1+e)}{R_I^2}\int_0^{l/2} dx'x' \int_0^{\infty} dv v^4\phi(v)\int_{0}^{1}d\chi\chi^2 = \frac{\pi(1+e)l^2}{12R_I^2}\int_0^\infty dv v^4\phi(v). \label{app:w_tot_transient2}
	\end{align}
	Similarly, we obtain 
	\begin{equation}
		\int_0^\infty d\mc{Y}\mc{Y}\int_{l}^{l+w} dx \int_{-\infty}^{\infty} d\vec v \phi(|\vec v|)\Theta(-\vec v\cdot \vec n(x))|\vec v\cdot \vec n(x)|\delta(\mc{Y}-\Delta \Om(x)) = \frac{\pi(1+e)w^2}{12R_I^2}\int_0^\infty dv v^4\phi(v).\label{app:w_tot_transient3}
	\end{equation}
	Equations~{(\ref{app:w_tot_transient1})},~{(\ref{app:w_tot_transient2})} and~{(\ref{app:w_tot_transient3})} lead to Eq.~{(\ref{eq:GM_first_irr})}.

\end{document}